\documentclass[twocolumn,pra,letterpaper,groupedaddress]{revtex4}

\usepackage{times,amsmath,amsfonts,amssymb,latexsym,textcomp}
\usepackage{color}
\usepackage{graphicx}
\usepackage{multirow}
\usepackage{bbm}
\usepackage{tabularx}
\usepackage{amsthm}
\usepackage{dsfont}
\usepackage{placeins}
\usepackage[lofdepth,lotdepth]{subfig}
\usepackage[T1]{fontenc}

\usepackage{float}
\usepackage{array}
\usepackage{caption}
\usepackage{braket}
\usepackage{MnSymbol}
\usepackage{tikz}
\usepackage[electronic]{ifsym}
\usepackage{booktabs}
\usepackage{subfig}

\renewcommand{\phi}{\varphi}

\renewcommand{\epsilon}{\varepsilon}

\begin{document}

\author{Mordecai Waegell$^{{a,b}}$, Eliahu Cohen$^{c,d}$, Avshalom Elitzur$^{a,d}$, Jeff Tollaksen$^{a,b}$, Yakir Aharonov$^{a,b,d,e}$}

\affiliation{\footnotesize{$^a$Institute for Quantum Studies, Chapman University, 1 University Dr., Orange, CA 92866, USA\\$^b$Schmid College of Science and Technology, Chapman University, 450 N Center St., Orange, CA 92866, USA\\$^c$Faculty of Engineering and the Institute of Nanotechnology and Advanced Materials, Bar Ilan University, Ramat Gan 5290002, Israel\\$^d$Iyar, The Israeli Institute for Advanced Research, POB 651 Zichron Ya'akov 3095303, Israel\\$^e$School of Physics and Astronomy, Tel Aviv University, Tel Aviv, Israel}}

\begin{abstract}
\textbf{Physical interpretations of the time-symmetric formulation of quantum mechanics, due to Aharonov, Bergmann, and Lebowitz are discussed in terms of weak values.  The most direct, yet somewhat naive, interpretation uses the time-symmetric formulation to assign eigenvalues to unmeasured observables of a system, which results in logical paradoxes, and no clear physical picture.  A \textit{top-down} ontological model is introduced that treats the weak values of observables as physically real during the time between pre- and post-selection (PPS), which avoids these paradoxes.  The generally delocalized rank-1 projectors of a quantum system describe its fundamental ontological elements, and the highest-rank projectors corresponding to individual localized objects describe an emergent particle model, with unusual particles, whose masses and energies may be negative or imaginary.  This retrocausal top-down model leads to an intuitive particle-based ontological picture, wherein weak measurements directly probe the properties of these exotic particles, which exist whether or not they are actually measured.\newline}

\footnotesize{\noindent Quantum Physics, Weak Values, Quantum Paradoxes, Time-Symmetry, Quantum Measurement}
\end{abstract}

\title{Quantum reality with negative-mass particles}
\maketitle

\section{Introduction}

In classical physics, knowing the initial state of a complete system is enough to infer any future state of that system, but due to the inherent indeterminism of measurement outcomes, this is not so in  quantum physics.  The time-symmetric, or two-state vector, reformulation of quantum mechanics was developed by Aharonov, Bergmann, and Lebowitz (ABL) \cite{aharonov1964time} in order to promote a description of quantum mechanics with the same level of completeness as classical physics.  For the quantum case, one needs not just the initial state of the system, but also its final state after a projective measurement, in order to make definite assertions about the quantum description of nature during the intervening time.  They developed the ABL formula, which provides the time-symmetric conditional probability to have obtained a given outcome for a strong projective measurement performed in the time between a particular initial state preparation and a particular final measurement outcome.  When these ABL probabilities are 1 or 0, they can be interpreted as describing simultaneous properties (eigenvalues) of the system between pre- and post-selection (PPS) --- even if no intermediate measurement is actually performed.  We call this the \textit{ABL interpretation}, and it gives rise to various logical PPS-paradoxes, which are directly connected to Kochen-Specker-type \cite{KS} quantum contextuality.  Because of these logical paradoxes, there is no consistent physical interpretation of such PPS situations, which has led to various \textit{ad hoc} explanations of their meaning, as in the cases of the 3-box paradox \cite{aharonov1991complete, ravon2007three}, the quantum Cheshire Cat \cite{aharonov2013quantum, denkmayr2014observation}, and the quantum pigeonhole effect \cite{aharonov2016quantum, waegell2018contextuality, waegell2017confined}, among others.

Some time later, Aharonov, Albert, and Vaidman (AAV) \cite{aharonov1988result,aharonov1990properties, aharonov1991complete} developed a new idea of what the quantum description should look like in the two-vector formalism for the situation where a very weak measurement (or no measurement at all) is made during the time between an initial preparation and a final measurement outcome, in terms of a quantity they called the \emph{weak value}, which is operationally more like an expectation value than an eigenvalue.  The weak value of an observable $\hat{A}$ is defined as $A_w = \langle \phi|\hat{A}|\psi\rangle / \langle \phi | \psi \rangle$.  Weak measurements were needed in the time-symmetric context in order to minimally disturb the pre- and post-selected states, thereby allowing us to answer questions which were not legitimate before. Analogously to {\it in vivo} studies, where biological experiments are performed on living organisms or cells, we would like to answer questions about freely-evolving quantum systems without perturbing them much. When the post-selection is not trivial, a strong measurement would alter either the pre- or post-selected state (or both), and hence we must employ weak measurements for keeping these quantum systems ``alive'' while being measured. This provides a new, richer description of quantum reality during the time interval between two projective measurements.  The weak value is generally complex, making its physical interpretation somewhat subtle \cite{Steinberg1995,Jozsa2007,Lobo2009,Cohen2018}, but unlike the ABL interpretation, the weak values do not produce any PPS-paradoxes, and can rather be seen as resolving the paradoxes of the ABL interpretation.  For a given PPS, the weak value of every observable property of the system is fully defined, and they collectively form a coherent quasi-classical picture of the system during the intervening time between the pre-selection and post-selection, which we call the \textit{weak value interpretation} (or weak reality) \cite{aharonov2018weak}.  As we will discuss, there is some physical motivation for this interpretation, since the weak value is observed in the limit of infinitesimally weak measurements with a PPS ensemble, and can thus be thought to be an existent property of the system, even when no actual measurement is performed.  In this sense, the weak value can be considered as more like an eigenvalue for a PPS, having a single definite value, but one that is operationally obscured by the spread of the pointer distribution in a weak measurement.  We call this interpretation quasi-classical because all of the weak values of different observables are mutually consistent and noncontextual, regardless of whether the observables commute.  Note that when we say the weak values provide a noncontextual value assignment, we mean only they are defined without reference to a measurement context.  This is unrelated to the fact that projector weak values prove Spekkens-type quantum contextuality when they are complex, or have real parts outside the range $[0,1]$ \cite{spekkens2005, pusey2014anomalous, piacentini2016experiment, kunjwal2018anomalous}.  

Both negative weak values and anomalously large weak values are ubiquitous in pre- and post-selected ensembles -- whenever $0<|\langle \phi|\psi\rangle|^2<1$, there is always a positive amplified weak value for the projector onto $\left(|\psi\rangle+|\phi\rangle\right)/\sqrt{2}$ and a negative weak value for the projector onto $\left(|\psi\rangle-|\phi\rangle\right)/\sqrt{2}$ (these values switch their roles if $-1<\langle \phi|\psi\rangle<0$).  Complex weak values are similarly commonplace.


Although lying outside the spectrum of the weakly measured operator $A$, weak values provide an effective description of the system between two strong measurements. This happens because the weak coupling to any operator $A$ of the pre- and post-selected system, through the interaction Hamiltonian, can be replaced by the c-number $A_w$ \cite{Aharonov2014,Vaidman2017}. This means, for instance, that in the examples studied below, a negative weak value of the projection operator onto some location would lead to an effective interaction term with inverse sign between the particle there and the weak probe. In particular, this implies the possibility of gravitational repulsion rather than attraction within the weak reality. Moreover, not only the gravitational mass, but also the inertial mass will be shown to admit a negative sign.

Finally, because the weak values at a given moment in time are defined using both the pre-selected state, $|\psi\rangle$, that propagates causally from past to future, and the post-selected state, $\langle\phi|$ that propagates retrocausally from future to past, the \emph{weak value interpretation} is also fundamentally nonclassical.


The weak reality leads to a \textit{top-down} \cite{aharonov2018completely} model of the physical reality during the time interval between a pre- and post-selection.  For an individual quantum particle, the model introduces additional positive-negative pairs (real or imaginary) of copies called \textit{counterparticles} which are generated by the quantum particle during pre-selection and absorbed during post-selection.  Each positive-negative pair has exactly opposite values for all physical properties, and thus their emergence from the vacuum obeys all conservation laws.  The counterparticles of a given system can interact with the counterparticles of other systems, but not with each other.

The ontology is called \textit{top-down} because it is the joint rank-1 projectors (outer products) of all $N$ quantum particles in an entangled state that are most fundamental, and the states of individual physical systems emerge only as sums over these objects.  The rank-1 projectors of an $N$-particle system correspond to delocalized $N$-point structures ($N$-structures), each made up of $N$ counterparticles of the same type - thus the $N$-structures come in the same four types, ($+1, -1, +i, -i$) as the counterparticles.  For the special case of an isolated non-entangled quantum particle, the 1-structures are just a single localized counterparticles.  Despite the retrocausal and top-down nature of the model, each of the counterparticles follows a continuous trajectory (world-line) through space-time, providing us with a clear (if exotic) physical picture of nature during the PPS interval.

We begin in Sec. \ref{PPS}, where we introduce the general hierarchy of all PPS paradoxes.  In Sec. \ref{counter}, we develop the top-down $N$-structure / counterparticle model of the weak value interpretation and discuss its properties.  In Sec. \ref{Examples}, we work through several significant examples to help develop a sense of the breadth of situations the new model can be applied to.  The quantum mirror is presented as a thought experiment to develop intuition about the ballistic properties of counterparticles (1-structures).  Several variations of the 3-box paradox are discussed, culminating with the Hardy paradox, \cite{hardy1992quantum, aharonov2002revisiting, lundeen2009experimental,yokota2009direct} which is reviewed as an example of top-down 2-structures and their properties, and a possible resolution of the paradox is discussed. Finally, we conclude with some closing remarks.

The supplemental information (SI) begins with a review of quantum measurement using a continuous pointer system, from weak measurement to strong projective measurement.  We also present a number of more general cases in the SI, which flesh out the weak reality picture, along with the analysis of other known PPS-paradoxes.  Lastly, the SI provides a generalization of top-down structures which allows them to produce the weak values for all measurement bases at once, thereby demonstrating that the weak reality is noncontextual.

\section{Pre-and-Post-Selection Paradoxes} \label{PPS}

There are a number of well-known PPS paradoxes, including the 3-box, the Quantum Cheshire Cat, and the Quantum Pigeonhole Effect, all of which are demonstrations of quantum contextuality \cite{KS,spekkens2005,leifer2005, tollaksen2007pre, pusey2014anomalous, pusey2015logical, waegell2018contextuality}.  We show here that all PPS paradoxes belong to a single general family of extended $N$-box paradoxes.  First we introduce some tools.  The weak value of any observable $A$ is given by the formula,
\begin{equation}
A_w \equiv \frac{\langle \phi |A | \psi\rangle}{\langle \phi | \psi\rangle},
\end{equation}
where the $|\psi\rangle$ is the pre-selection and $|\phi\rangle$ is the post-selection.  


Noting the spectral decomposition $A = \sum_i \lambda_i \Pi_i$, where $\Pi_i \equiv |a_i\rangle\langle a_i |$ is the projector onto a given eigenstate and $\lambda_i$ is the corresponding eigenvalue, it is clear that $A_w = \sum_i \lambda_i (\Pi_i)_w$.  

The ABL probability formula gives the conditional probability to obtain a particular measurement result if a projective measurement was made during the time between the pre-selection of $|\psi\rangle$ and the post-selection of $|\phi\rangle$.  The outcomes of a projective measurement of observable $A$ are the projectors $\Pi_i$, and the formula is,

\begin{equation}
P_{\textrm{ABL}}(\Pi_i=1 \space \; \vert \; \psi, \phi, \mathcal{B})  = \frac{  |\langle \phi | \Pi_i |\psi\rangle|^2  } { \sum_{k\in\mathcal{B}} | \langle \phi | \Pi_k |\psi\rangle|^2 } = \frac{  |(\Pi_{i})_w|^2  } { \sum_{k\in\mathcal{B}} | (\Pi_{k})_w|^2 }. \label{ABL}
\end{equation}

This formula shows that the key to understanding the general family of PPS paradoxes lies in the weak values of the projectors $\{\Pi_i\}$ onto the measurement basis $\mathcal{B}$.  Of special importance is the case where the measurement basis has just two possible outcomes (dichotomic).  In this case, if either projector has weak value 1 (0), then the ABL probability to obtain that outcome for an intermediate projective measurement is also 1 (0).  This follows from Eq .\ref{ABL}, since $\sum_{k\in\mathcal{B}} (\Pi_k)_w = 1$ means the weak values of the two projectors must be 1 and 0 in either case.

To obtain a PPS paradox, we must consider  multiple different coarse-grained dichotomic measurement bases $\mathcal{B}_n$ where the lower-rank projectors from the eigenbasis of $A$ are combined in different ways into higher-rank projectors of the coarse-grained bases.  A PPS paradox requires that there are multiple coarse-grained dichotomic bases with weak values 1 and 0, and there is a logical contradiction between all of the different projectors with ABL probability 1.  Note that all PPS paradoxes require some projector weak values with negative real parts \cite{cohen2015voices, elitzur20161}, which demonstrates the connection between the proofs of Kochen-Specker contextuality in the paradoxes \cite{KS,waegell2017confined,waegell2018contextuality}, Spekkens contextuality in the anomalous weak values \cite{spekkens2005,leifer2005,pusey2014anomalous, pusey2015logical,kunjwal2018anomalous}, and the Kirkwood-Dirac distribution \cite{dressel2015weak}.

The simplest example is the 3-box paradox \cite{aharonov1991complete, ravon2007three}, which is a 3-level system with weak values $(\Pi_1)_w = 1$, $(\Pi_2)_w = 1$, and $(\Pi_3)_w = -1$.  We construct two coarse-grained dichotomic measurement bases, $\mathcal{B}_1 = \big(\Pi_1 + \Pi_3, \Pi_2 \big)$ and $\mathcal{B}_2 = \big(\Pi_1, \Pi_2 + \Pi_3\big)$, both of which have projectors with weak values 1 and 0.  From $\mathcal{B}_1$, the ABL rule tells us that the system must be in state $\Pi_2$, while from $\mathcal{B}_2$ it tells us that it must be in the orthogonal state $\Pi_1$.  This logical contradiction is the PPS paradox.

All PPS paradoxes are of this kind, which makes the identification of the complete family fairly straightforward.  The key was the set of weak values $(1,1,-1)$ in the fine-grained measurement basis $\mathcal{B}_0$, and there is a similar key paradox for each number $N$ of elements in $\mathcal{B}_0$, which is given by $(1,...,1,-1)/(N-2)$.  These are the fundamental $N$-box paradoxes, but each one can be embedded into a higher-dimensional system, by setting all additional projector weak values to zero, (e.g. $(1,1,-1,0)$).  The $N$-box paradoxes requires $N-1$ different coarse-grained bases, for which the ABL rule ultimately shows that the system cannot be in any of its states.  The Quantum Cheshire Cat and Quantum Pigeonhole Effect are both examples of the 4-box paradox, but with projectors of different ranks.

From these key paradoxes, an infinite extended family can be generated in higher dimensions by taking any weak value in the set and breaking it into a sum of two or more weak values.  For example, this can be done by breaking a weak value 1 into two 1/2s, or by breaking a 0 into a 1 and -1.  In all of these cases, the fundamental logic of the PPS paradox derives from the underlying key $N$-box paradox.  The family of box paradoxes in \cite{aharonov1991complete} all have the 3-box paradox as their key, and are obtained by breaking the -1 into a $-n$ and $n-1$ 1s, where $n\geq 2$.  

We can always find a PPS for any possible set of complex projector weak values $w_i$ that obey $\sum_i w_i = 1$ by taking pre-selection $|\psi\rangle = \sum_i w_i |a_i\rangle$ and post-selection $|\phi\rangle = \sum_i |a_i\rangle$ (regardless of normalization), resulting in weak values $(\Pi_i)_w = w_i$, so all such cases can be realized.  This can be generalized to the infinite PPS class $|\psi\rangle = \sum_i (w_i c_i) |a_i\rangle$ and  $|\phi\rangle = \sum_i (1/c_i^*)|a_i\rangle$, which also produce the weak values $(\Pi_i)_w = w_i$, for any set of complex coefficients $\{c_i\}$ --- and there are many other choices of PPS that will also produce these weak values.

The physical interpretation of a key PPS paradox depends strongly on what quantum system is used to implement it, and the physical manifestation of the paradox is different in each case.  This has led to many counterintuitive physical examples.  It also provides a road map to search for new implementations using the key paradoxes as a mathematical backbone. 

Of course, in the weak value interpretation, there are no logical contradictions among the weak values, although they are not generally eigenvalues.  But in this interpretation, it is the projectors in the finest-grained basis that are most fundamental, which are formally states of every system in the universe -- or at least every system in an entangled state.  Thus, for two entangled electron spins, the weak values of four rank-1 projectors in any 2-spin basis are well-defined, but attempts to deduce the individual weak states of the spins may result in a logical contradiction, as in the 4-box paradox, and many other cases.  The counterparticle model provides a straightforward quasi-classical description of the weak values in these cases.

\section{Counterparticles and Top-Down $N$-structures}  \label{counter}

In the weak reality picture, we replace the conventional wave interference picture of quantum mechanics with a particle scattering picture (see also \cite{aharonov2018weak, aharonov2018extraordinary}).  Instead of the entangled superposition state of the system and pointer, collapse, and interference, the pointer undergoes a single quasi-classical impulsive interactions corresponding to all of the other $N$-structures it interacts with during the PPS interval - but this is truly only in the singular limit when there is no interaction at all.  To first order in $d/\epsilon$, this gives physical predictions identical to the usual entanglement-based treatment.  

The new model requires the introduction of new counterpart particles to the original (\textit{counterparticles}), which emerge in positive-negative pairs, and which are linked in a corresponding top-down $N$-structure, which is represented as a graph on $N$ vertices.  The new structure of counterparticles is emitted from the original particles during the pre-selection process and then reabsorbed by them during the post-selection process, such that the original particles collect the measurement back-action from each $N$-structure's interactions with the pointer.  An entire $N$-structure may be positive or negative, and real or imaginary, and it contains $N$ identical counterparticles at its vertices - and thus a 1-structure is just a single counterparticle.  We thus have three new types of particles in our time-symmetric ontology, for total of four types: $(+1, -1, +i, -i)$.  In summary, this time-symmetric model tells an explicit quasi-classical story of (counter)particles that move on definite trajectories through space-time, and the top-down structures that connect sets of entangled particles.


For a given rank-1 projector of an entangled $N$-body state, with weak value $x+iy$, there are, on average, $|x|$ $N$-structures composed of real counterparticles of type sign$(x)$, and, on average, $|y|$ $N$-structures composed of imaginary counterparticles of type $i$sign$(y)$, with the $N$ counterparticles in each structure spanning $N$ sites/states of the rank-1 projector.  On any given run there will always be an integer number of particles of various types, forming complete integer $N$-structures, but the actual probability distributions for $x$ and $y$ are irrelevant for the model, since we are at the limit where there is no back action on a measurement device at all.

Each $N$-structure is a connected graph, but in general it is not fully connected and to find the actual edges, we first define 2-structures for all pairs of subsystems by summing the rank-1 projectors over all other systems, and then keep only the edges from these different pairwise 2-structures for the full $N$-structure.  Once all $N$-structures are properly defined, they cleanly encode the system's weak values for all projectors of all ranks.  Specifically, $N$-structures for higher-rank projectors corresponding to $n<N$ subsystems are simply the subgraphs on the $n$ corresponding vertices of the $N$-structure.

If the momentum of a pointer system couples to a quantum system at just one location (the highest-rank projector of an entangled system) during the time between pre- and post-selection, the pointer receives a quasi-classical impulse (instead of entering an entangled superposition) corresponding to the sum of the counterparticles at that location.  We begin with a few single-particle examples in order to highlight the simple ballistic properties of counterparticles during local interactions, and then we consider the Hardy paradox as an example with 2-structures.

Individual local measurements during a PPS reveal nothing about the top-down structure.  Measurements of the lower-rank projectors which form this structure are discussed alongside an example of 3 entangled 2-level systems in the SI.

\section{Illustrative, Paradoxical, and Counterintuitive Examples} \label{Examples}

\subsection{2-level Systems}

First, consider the example of a single particle that is both pre- and post-selected in a superposition of two boxes, $|\psi\rangle = |\phi\rangle = \big(|1\rangle  + |2\rangle\big)/\sqrt{2},$ resulting in weak values $(\Pi_1)_w = (|1\rangle\langle 1|)_w = 1/2$ and $(\Pi_2)_w = (|2\rangle\langle 2|)_w = 1/2$.  In the weak value ontology, the simplest possible distribution has the particle in each box with probability 1/2.  To be clear, there are more complicated distributions involving additional positive-negative pairs of counterparticles which will produce the same weak values, and freely switching between any such distribution is a symmetry of the interpretation.

Next, consider an example of a 2-level system, with orthogonal projectors $\Pi_1$ and $\Pi_2$, having weak values  $(\Pi_1)_w = -1/3 + 3i/2$ and $(\Pi_2)_w = 4/3 - 3i/2$ in a given PPS (there are infinitely many possible PPS choices that produce these values).  We can work out the probability distributions of the real counterparticles and imaginary counterparticles separately, since the positive-negative pairs are independent for the two cases.  Starting with the real part, the simplest distribution is that with probability 2/3 there is a positive particle in $\Pi_1$ and nothing in $\Pi_2$, and with probability 1/3 there are two positive particles in $\Pi_1$ and a negative one in $\Pi_2$.  Then the simplest independent probability distribution for the imaginary part has one positive particle in $\Pi_1$, and a negative one in $\Pi_2$ with probability 1/2, and also with probability 1/2, two positive particles in $\Pi_1$, and two negative particles in $\Pi_2$.

\subsection{The Quantum Mirror}

When a pointer interacts with a system in only one location, the impulse is due simply to the sum of impulses from the individual counterparticles at that location, which provides a very simple quasi-classical ballistic picture of these types of interactions during the PPS.

Consider a Mach-Zehnder Interferometer (MZI) through which a single particle passes.  On arm II of the MZI we have the usual macroscopic mirror, but on arm I we instead have a tiny mirror of mass $m$, which we will approximate as a single coherent superposition state with a small position uncertainty $\Delta x$, and a large momentum uncertainty $\Delta p$.  We presume that the mirror still scatters the particle in the same direction as a macroscopic mirror, and that $m$ is sufficiently large compared to the particle's momentum so that the energy of the particle is not changed significantly.  Furthermore, the particle must have an energy uncertainty, $\sigma$, which is large enough that this change in energy does not significantly reduce the visibility of interference at the second beamsplitter.

Ignoring the back-action on the particle's energy, the coupling Hamiltonian is $\hat{H} = g(t)\hat{\Pi}^p_\textrm{I} \hat{X}^m$, where $g(t)$ is a time-dependent coupling strength, which integrates to $s = \int_\textrm{int} g(t) dt$ during the scattering interaction.  This is the interaction for a measurement of $\hat{\Pi}^p_\textrm{I}$ using the momentum of the quantum mirror as the pointer system (see the SI).  We are interested here in the limit $s \ll \Delta p$ where this becomes a weak measurement of the projector $\hat{\Pi}^p_\textrm{I}$ (see also \cite{aharonov2013classical,dziewior2019universality}).

Before the particle in the MZI reaches the quantum mirror, the product state of the two systems is,
\begin{equation}
\int |dp\rangle \tilde{\varphi}(p)\big(|\textrm{I}^p\rangle + |\textrm{II}^p\rangle\big)  /\sqrt{2},
\end{equation}
where $\tilde{\varphi}(p)$ is the momentum-space wavefunction of the mirror, and $|\psi^p\rangle = |\textrm{I}^p\rangle + |\textrm{II}^p\rangle$ is the particle's pre-selected state.  After the particle scatters off of the quantum mirror, the two evolve into the entangled state,
\begin{equation}
\int |dp\rangle\big(\tilde{\varphi}(p - s)|\textrm{I}^p\rangle + \tilde{\varphi}(p)|\textrm{II}^p\rangle\big)  /\sqrt{2}.
\end{equation}
Now, to allow a general post-selection $|\phi^p\rangle$, we must allow that an arbitrary phase shifter be placed in one arm of the MZI, and that the second beamsplitter can be chosen with any transmissivity.  Whatever the choice, the final state of the mirror will be $\tilde{\varphi}\big(p - s(\Pi^p_\textrm{I})_w\big)$ (see SI).

\begin{figure}[t!]
    \centering
    \includegraphics[width=3.2in]{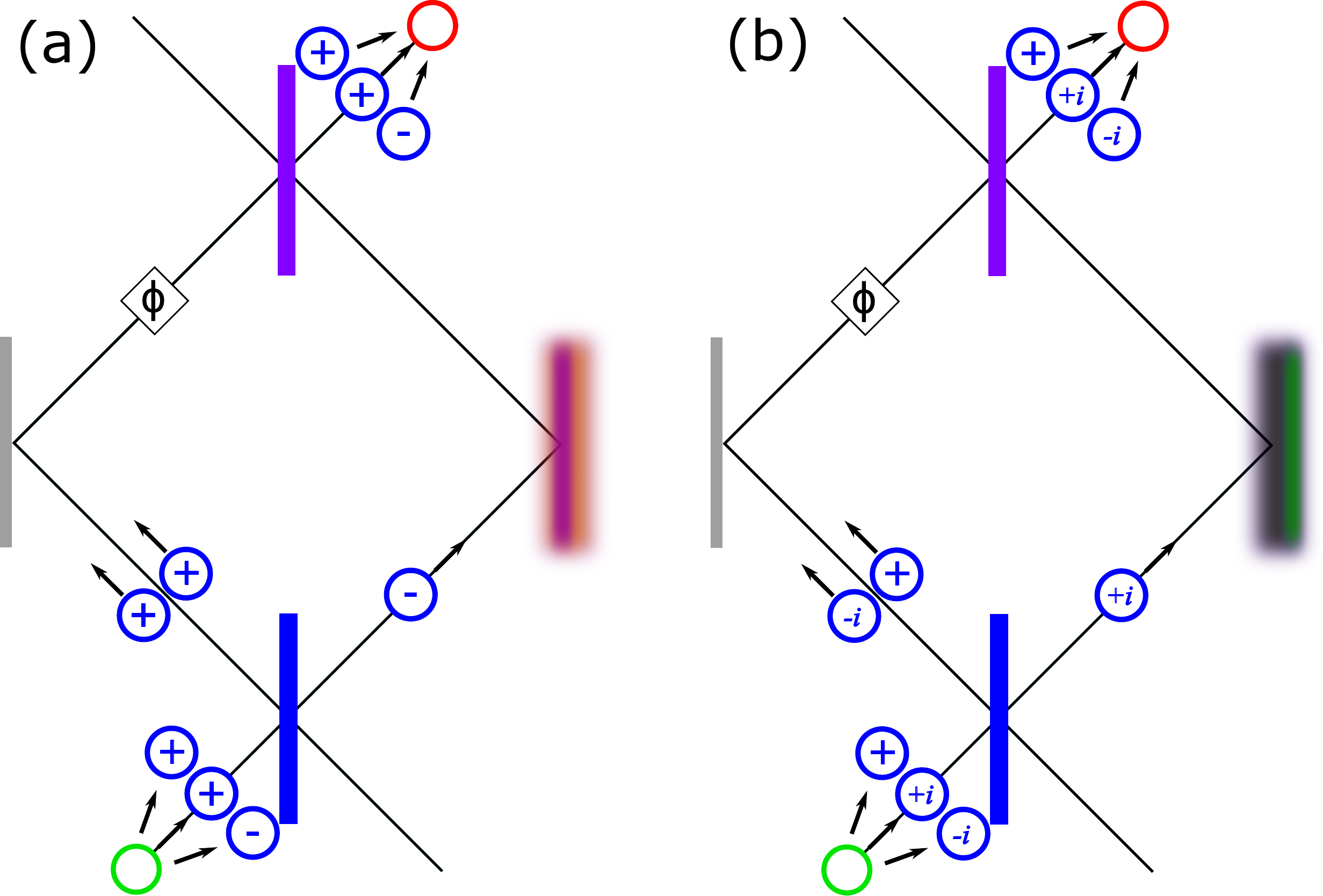}
    \caption{The quantum mirror for pre-selection, $|\psi^p\rangle = |\textrm{I}^p\rangle + |\textrm{II}^p\rangle$, and two different post-selections, corresponding to different choices of transmissivity in the second (purple) beamsplitter and different setting of the phase shifter. (a) Counterparticle configuration for post-selection, $|\phi^p_1\rangle = -|\textrm{I}^p\rangle + 2|\textrm{II}^p\rangle$.  During pre-selection, the particle emits a positive-negative pair of counterparticles, and the negative one scatters off the quantum mirror, whose momentum wavefunction is depicted in orange.  The counterparticles are reabsorbed during the post-selection, and the post-selected quantum mirror receives a negative impulse, as shown by the magenta sub-ensemble.   (b) Counterparticle configuration for post-selection, $|\phi^p_2\rangle = i|\textrm{I}^p\rangle + (1-i)|\textrm{II}^p\rangle$.  During pre-selection, the particle emits an imaginary positive-negative pair of counterparticles, and the imaginary positive one scatters off the quantum mirror, whose position wavefunction is depicted in black.  The counterparticles are reabsorbed during the post-selection, and the post-selected quantum mirror receives is translated in the positive direction, as shown by the green sub-ensemble.}
    \label{Mirror}
\end{figure}

We consider two different cases, and because they have no affect on the weak values, we do not bother with normalization factors.  First, let the post-selection be $|\phi^p_1\rangle = -|\textrm{I}^p\rangle + 2|\textrm{II}^p\rangle$, which results in the weak values $(\Pi^p_\textrm{I})_w = -1$ and $(\Pi^p_\textrm{II})_w = 2$.  In this case, the quantum mirror has been given a negative impulse, and has been pulled inward rather than pushed outward by the scattering of the particle.  This counterintuitive effect is a straightforward prediction of quantum theory for a coherent quantum mirror.

The counterparticle representation of this case is quite intuitive, in that it says that there were two real positive particles on arm II, and a negative real particle that bounced off of the mirror in arm I.  This particle has negative mass, and thus negative momentum relative to its direction of propagation, and so the impulse to reflect it is also negative, pulling the mirror inward. 


Next, let the post-selection be $|\phi^p_2\rangle = i|\textrm{I}^p\rangle + (1-i)|\textrm{II}^p\rangle$, which results in weak values $(\Pi^p_\textrm{I})_w = i$ and $(\Pi^p_\textrm{II})_w = 1-i$.  In this case, the mirror has received zero impulse, but it has undergone a translation in space.  This is also a straightforward prediction of quantum theory for a coherent quantum mirror.

Again, the counterparticle representation provides a clear description, where there is only a positive imaginary particle on arm I, and on arm II there is a positive real particle and a negative imaginary particle.  When the imaginary particle scatters off the quantum mirror, the mirror exerts an impulse on the imaginary particle, and in reaction, the imaginary particle translates the mirror with no impulse.  This case serves as a guide for how we should conceptualize scattering interactions involving the imaginary counterparticles in this quasi-classical picture.

These examples help to build intuition for this picture in the regime of observable weak measurements, but our general focus in this article in on the underlying ontology that is present during a PPS when no intermediate interaction occurs at all.

\subsection{The 3-box Paradox}

\begin{figure}
    \centering
    \includegraphics[width=3.2in]{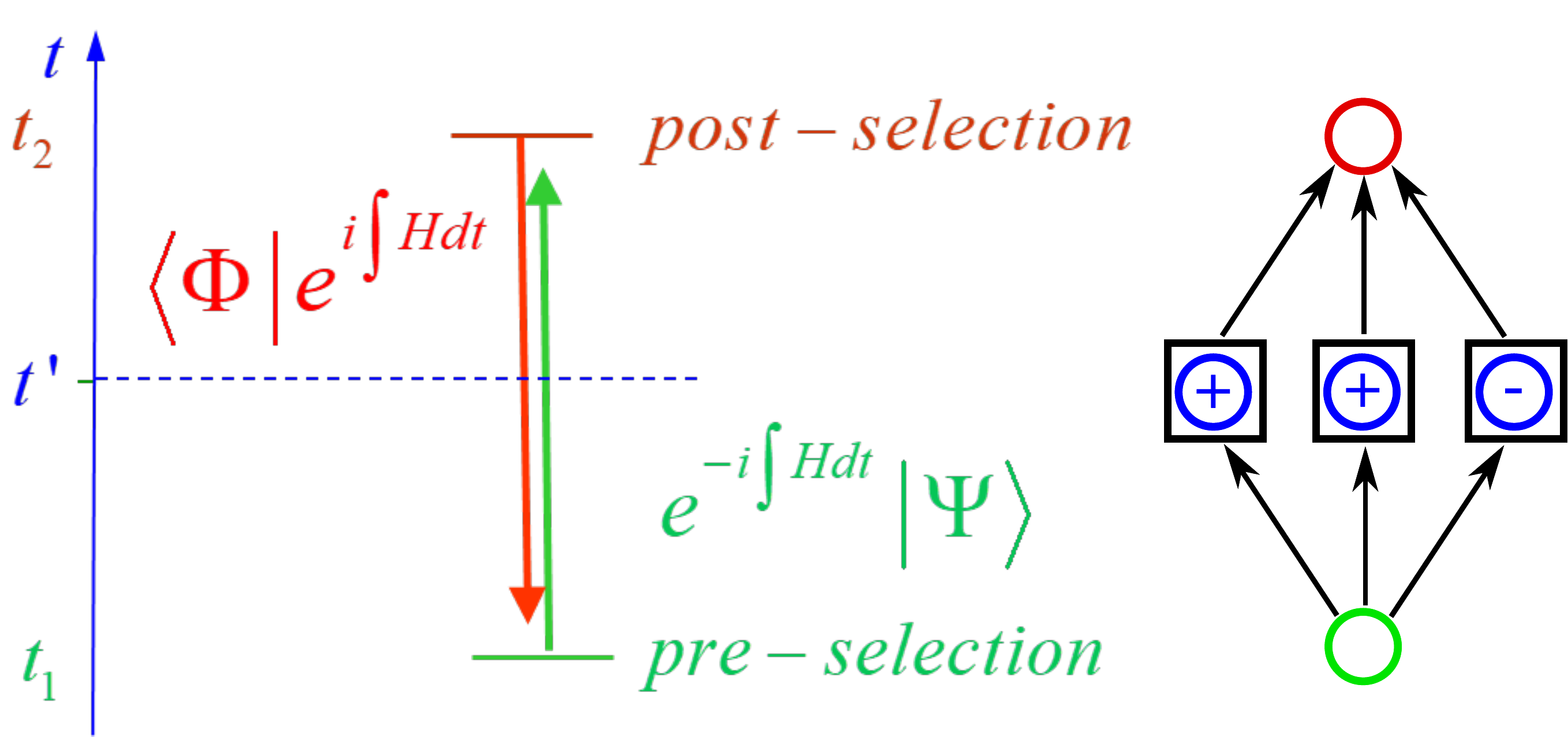}
    \caption{The 3-box paradox as explained in the counterparticle model.  An extra positive-negative pair of copies of the original particle is created during the pre-selection and reabsorbed during the post-selection.  In general, the configuration of counterparticles may change dynamically due to unitary evolution during the PPS, as in the case of the Disappearing and Reappearing particle.}
    \label{3box}
\end{figure}

The 3-box paradox is the simplest case \cite{aharonov1991complete, ravon2007three}.  For projector weak values $(1,1,-1)$ there is a real positive particle in the first state, another in the second, and a real negative particle in the third.  This example was the primary motivation for the development of the counterparticle model.  Fig. \ref{3box} illustrates how the extra positive-negative pair is created during the pre-selection and reabosrbed during the post-selectin.

Next, we will discuss several other examples for which the 3-box is the key paradox.

\subsection{Vaidman's Nested Interferometer Paradox}

\begin{figure}[h]
     \centering
     \includegraphics[width = 2.8in]{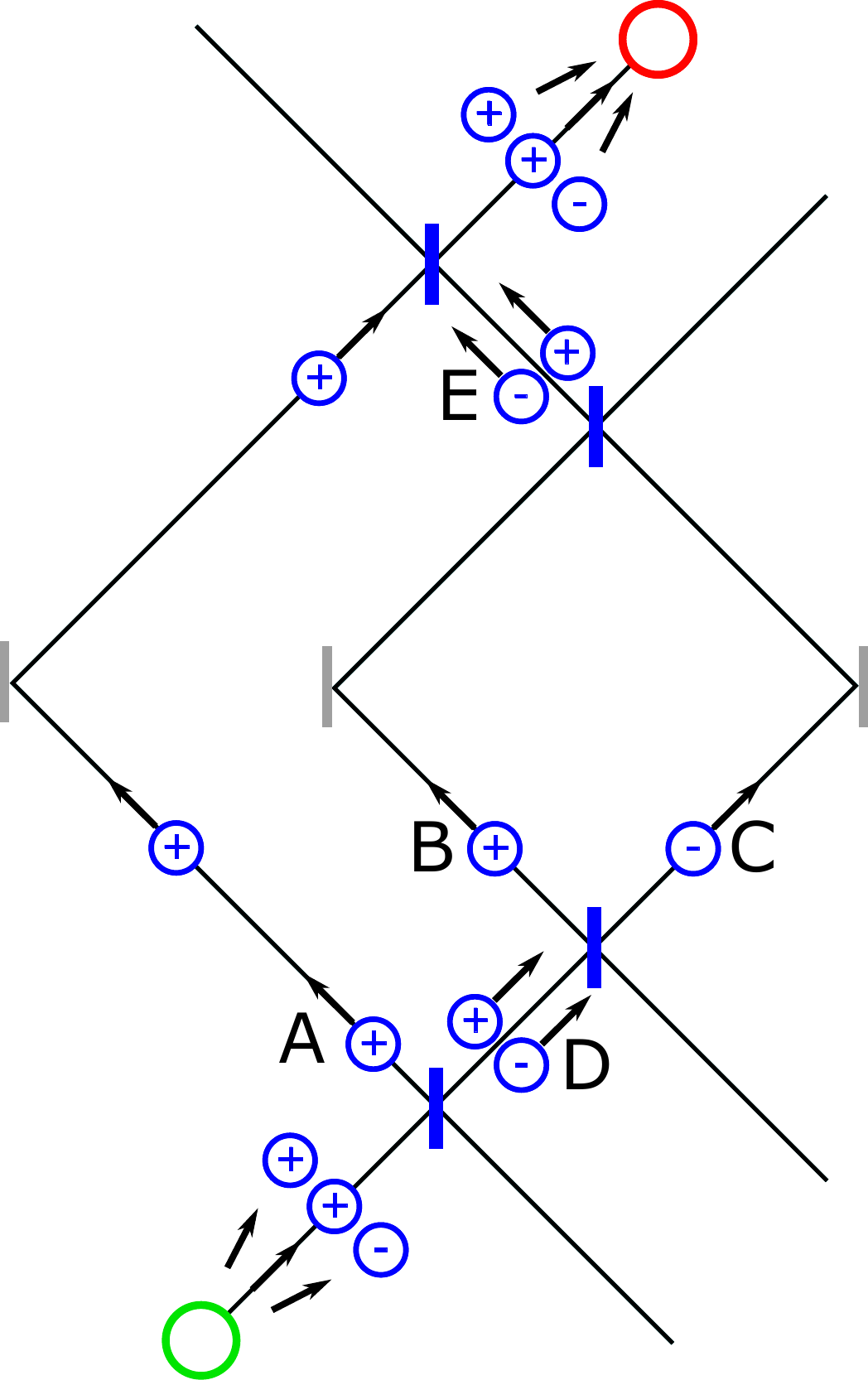}
     \caption{The counterparticle configuration for Vaidman's nested Mach-Zehnder Interferometer experiment.  During the pre-selection, the photon emits a positive-negative pair of counterparticles.  A positive-negative pair move together in $D$ and $E$, and thus a weak measurement will detect zero particles in these regions.  After the second beamsplitter, weak measurements detect the positive particles in $A$ and $B$, and the negative particle in $C$. }
     \label{Nested}
 \end{figure}

 Let us consider Vaidman's experiment \cite{vaidman2013past, aharonov2018weak} with a nested pair of MZIs in which, given the post-selection, there is a weak trace in both arms of the inner interferometer, even though there is no trace in the only paths into, or out of, the inner interferometer.  Let $A$ be the arm of the outer interferometer that bypasses the inner interferometer, $D$ be the arm that leads into the inner interferometer, $E$ the same arm leading back out, and $B$ and $C$ the arms of the inner interferometer --- as shown in Fig. \ref{Nested}
 
At time $t_1$, after the particle has entered the outer interferometer, but before it would reach the inner interferometer, the weak values of the relevant projectors are $|A|_w = 1$, $|D|_w =0$ --- thus there is no weak trace leading into the inner interferometer.  At time $t_2$, when the particle would be inside the inner interferometer, the weak values of the relevant projectors are $|A|_w = 1$, $|B|_w = 1$, and $|C|_w = -1$, and thus there is a weak trace in both arms of the inner interferometer.  Finally at time $t_3$, after the particle would exit the inner interferometer, the weak values of the relevant projectors are $|A|_w = 1$, $|E|_w =0$, and there is no weak trace leading out of the interferometer.

Thus, at $t_2$ we have obtained exactly the 3-box paradox, with a real positive particle in arm $A$, another in arm $B$, and a real negative particle in arm $C$.  Now, since the positive-negative counterparticle pair must be emitted during the pre-selection, and reabsorbed during the post-selection, and each must follow a complete world-line through the PPS, we interpret $|D|_w =0$ and $|E|_w =0$ as the positive-negative counterparticle pair from $B$ and $C$ moving together, so that they apparently mask one another. However, there are operators whose weak values are not zero in arms $D$ and $E$. If we describe the states at arms $B$ and $C$ as the two eigenstates of $\sigma_z$, then $\sigma_z$ has a non-zero weak value within the inner interferometer, but not in arms $D$ and $E$. Nevertheless, the weak value of $\sigma_x$ (or $\sigma_y$) which corresponds to a flip from arm $B$ to arm $C$ (up to a phase) does not vanish in arms $D$ and $E$. These two operators, which are sensitive to the relative phase between the arms, are nonlocal and cannot be instantaneously measured in $D$ and $E$, but thanks to the Heisenberg equations they change in time into locally measurable operators.  


It is interesting to note the momentum exchange of the particle-counterparticle pair with the inner beamsplitters in the case that the latter are heavy but not held fixed. The net momentum at the entrance to the inner interferometer is zero, because the particle and counterparticle have opposite momenta. However, within the interferometer they gain a net momentum to the left, which means that the first inner beamsplitter must receive a net momentum to the right in order to conserve the total momentum. Also the second inner beamsplitter must receive from the pair a net momentum to the right. The proposed ontology can therefore give rise to new predictions which were not obtainable before. We can see this way again that the arms $D$ and $E$ are not empty but rather contain a particle-counterparticle pair.   

In the ABL interpretation, and noting that $|B|_w + |C|_w = 0$, we can conclude that the particle is definitely in path $A$ at all three times $t_1$, $t_2$, and $t_3$, and that it was never in the inner interferometer at all, nor in the paths leading into or out of it.  However, if at $t_2$ a measurement were performed in the basis $\{A,B,C\}$ the ABL formula predicts probability $1/3$ to find the particle in $B$, and probability $1/3$ to find it in $C$.  The appearance of the particle in the inner interferometer, despite it having zero probability to be found entering or exiting, is then the same sort of paradox as the appearing and disappearing particle \cite{aharonov2017case}.

\subsection{The Quantum Cheshire Cat as a 3-box Paradox}

\begin{figure}
    \centering
    \includegraphics[width=2in]{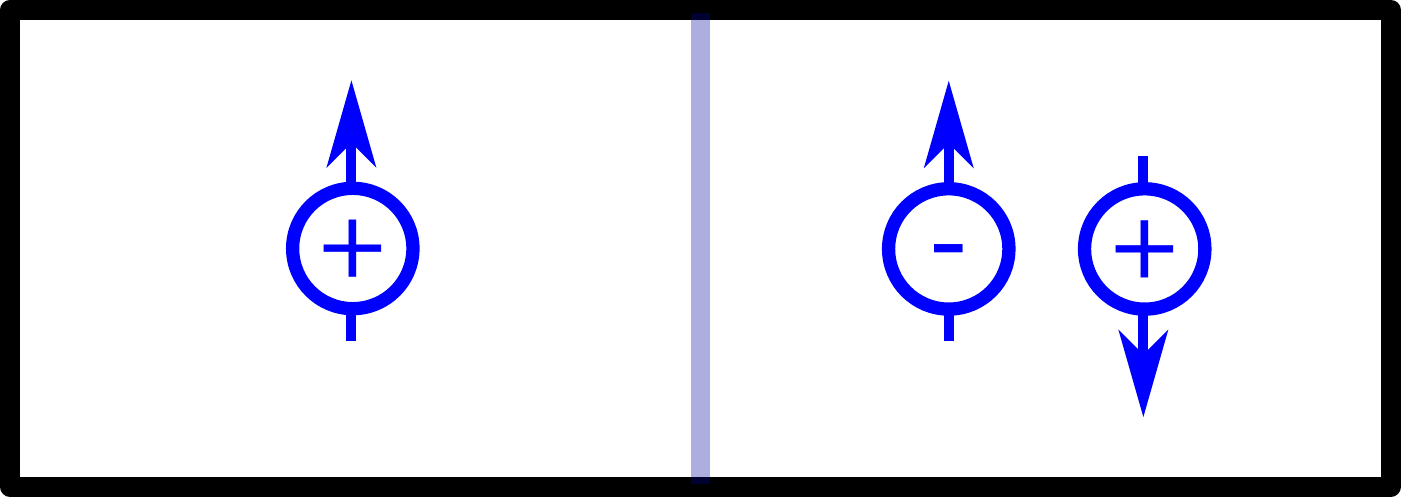}
    \caption{The three particles that are present during in the 3-box-paradox version of the Quantum Cheshire Cat.  The net mass on the right side is 0, but the net spin is twice the spin eigenvalue, creating the appearance of a disembodied anomalous spin.}
    \label{fig:my_label}
\end{figure}

Consider a spin in cavity with a spin-sensitive mirror in the center.  The spin begins on the left in the state $2|L\uparrow\rangle + |L\downarrow\rangle$.  The mirror is transparent to $|\downarrow\rangle$, and acts as a 50/50 beam splitter for $|\uparrow\rangle$, and thus after striking the mirror, the state has evolved to $|\psi\rangle = |L\uparrow\rangle + |R\uparrow\rangle + |R\downarrow\rangle$.  After a second pass through the mirror, the state is measured as $2|R\uparrow\rangle + |L\downarrow\rangle$.  Counterpropagating this through the mirror gives us $|\phi\rangle = |L\uparrow\rangle - |R\uparrow\rangle + |R\downarrow\rangle$.  $|\psi\rangle$ and $|\phi\rangle$, are, respectively, the pre- and post-selection during the time between the first and second pass of the mirror.  The weak values of the four rank-1 projectors are then,
\begin{equation}
    \begin{array}{cccc}
        |L\uparrow|_w = 1, & |L\downarrow|_w = 0, &|R\uparrow|_w = -1, & |R\downarrow|_w = 1, \\
    \end{array}
\end{equation}
and for the rank-2 projectors they are,
\begin{equation}
    \begin{array}{cc}
        |L|_w = |L\uparrow|_w + |L\downarrow|_w = 1, & |R|_w = |R\uparrow|_w + |R\downarrow|_w = 0, \\
    \end{array}
\end{equation}
and
\begin{equation}
    \begin{array}{cc}
        |\uparrow|_w = |L\uparrow|_w + |R\uparrow|_w = 0, & |\downarrow|_w = |L\downarrow|_w + |R\downarrow|_w = 1. \\
    \end{array}
\end{equation}

From the rank-2 projector weak values, we would conclude that the particle must be on the left ($|L|_w=1$), with spin down ($|\downarrow|_w=1$), but this directly contradicts the rank-1 projector weak values ($|L\downarrow|_w=0$).  Furthermore, a weak measurement of the mass on the right arm will detect nothing, while a weak measurement of the spin on the right arm will detect twice the expected value down, so we seem to have a disembodied spin, as in the original Quantum Cheshire Cat.

The counterparticle description is quite straightforward.  There is a standard spin-up particle on the left, a standard spin-down particle on the right, and a negative spin-up particle on the right - which produces the same magnetic field as a positive spin-down particle, due to its opposite charge.  Instead of a disembodied spin on the right, we have two particles whose masses sum to zero, but whose magnetic fields add constructively (and to an anomalously large value).

Interestingly, there appears to be a way to experimentally validate this description.  The two counterparticles in the right side of the cavity have opposite mass and charge, and are in opposite spin states.  The opposite charge and spin result in identical magnetic moments, which is why we observe zero mass and charge (if they are charged particles), but nonzero magnetic moment.  Now, if an electric field were turned on within the right cavity, the particles would experience opposite Coulomb force, but they also have opposite inertial mass, and their accelerations are identical.  However, if a magnetic field gradient were turned on instead, they would have opposite acceleration, because they would experience identical magnetic force.  

Applying such a magnetic field would allow us to separate the two counterparticles within the right side of the cavity, and then their individual masses and magnetic moments could be weakly measured, before reversing the direction of the magnetic field to put the particles back together, such that the post-selection is not disturbed.  Thus, a suitably modified experiment could show that what appears to be a spin without a mass is really two counterparticles partially masking each other.

Note that the inertial mass of counterparticles on the right side is indeed negative since the effective momentum associated with these particles is negative (the weak value of the corresponding projector is negative), but their velocity is positive (it was not pre- and post-selected at all).

\subsection{The Hardy Paradox}

Hardy's paradox \cite{hardy1992quantum, aharonov2002revisiting} has been widely studied in connection with quantum nonlocality.  It features two MZIs, one traversed by an electron and the other by a positron.  There is a point of intersection between an arm of one interferometer and an arm of the other, such that if the electron and positron both travel down that arm, they are annihilated and two gamma-ray photons are produced.  This creates effectively a 5-level system --- the positron in one of two arms, together with the electron in one of two arms give four states, and the photons produced by the annihilation are the fifth.

Let the arms of the interferometer traversed by the positron be $|L^+\rangle$ and $|R^+\rangle$, and the arms of the interferometer traversed by the electron be $|L^-\rangle$ and $|R^-\rangle$, with the arms $|R^+\rangle$ and $|L^-\rangle$ intersecting such that $|R^+\rangle|L^-\rangle \rightarrow |\gamma\rangle|\gamma\rangle$.  The interferometers have bright ports $b^\pm$ and dark ports $d^\pm$ and are aligned such that if the two interferometers were moved apart so that there could be annihilation, then no particles would be detected at either dark port $d^\pm$.  For the present configuration, the possibility of annihilation alters the wavefunction such that there is a 1/16 probability to detect both the electron and the positron at their dark ports, which can be interpreted as indicating that the two particles detected one another without ever interacting (i.e., interaction-free measurement \cite{elitzur1993quantum}).
\begin{figure}[t!]
    \centering
    \includegraphics[width=3.2in]{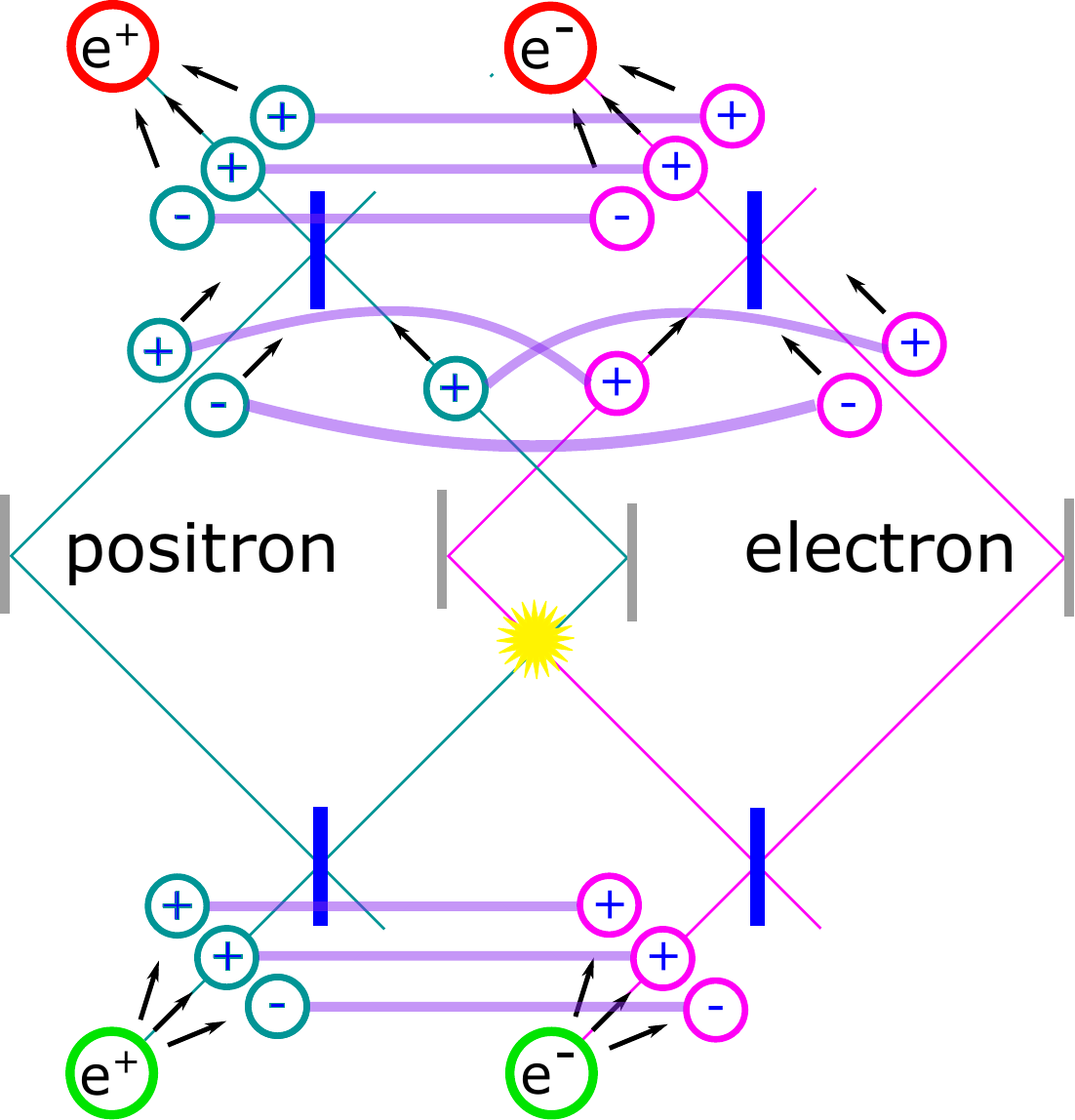}
    \caption{An illustration of the top-down 2-structures and their counterparticles as they pass through the overlapping MZIs in the Hardy Paradox.}\label{Hardy}
\end{figure}
After passing through the first beamsplitter of their MZIs, the positron is in the state $(|L^+\rangle + i|R^+\rangle)/\sqrt{2}$ and the electron is in the state $(|R^-\rangle + i|L^-\rangle)/\sqrt{2}$ which result in the joint product state,
\begin{equation}
|\psi_0\rangle = ( |L^+\rangle|R^-\rangle   + i  |L^+\rangle|L^-\rangle + i  |R^+\rangle|R^-\rangle   -|R^+\rangle|L^-\rangle)/2.
\end{equation}
Once the two pass the intersection point, this evolves into the entangled state,
\begin{equation}
|\psi\rangle = ( |L^+\rangle|R^-\rangle   + i  |L^+\rangle|L^-\rangle + i  |R^+\rangle|R^-\rangle   -|\gamma\rangle|\gamma\rangle)/2,
\end{equation}
which is to say that the possibility of annihilation has created an entanglement correlation between the electron, positron, and gamma-ray photons.  This entangled state is the pre-selection we will consider here.

For the post-selection, we of course take the paradoxical case where the detectors at both dark ports d$^\pm$ have clicked.  After propagating them retrocausally back through the second beamsplitter of their MZIs, the positron is post-selected in the sate $(|L^+\rangle - i|R^+\rangle)/\sqrt{2}$ and the electron in the state $(|R^-\rangle - i|L^-\rangle)/\sqrt{2}$, resulting in the product state,
\begin{equation}
|\phi\rangle = ( |L^+\rangle|R^-\rangle   - i  |L^+\rangle|L^-\rangle - i  |R^+\rangle|R^-\rangle   - |R^+\rangle|L^-\rangle)/2.
\end{equation}
With this pre- and post-selection, both valid during the time interval between the possible annihilation event and the arrival of the particles at the second beamsplitter of each interferometer, we construct the weak values $|L^+R^-|_w = -1$, $|L^+L^-|_w = 1$, $|R^+R^-|_w = 1$, $|R^+L^-|_w = 0$, and $|\gamma \gamma|_w = 0$, using compact notation $|R^+L^-| \equiv |R^+L^-\rangle\langle R^+L^-|$, which correspond to the 3 (nonzero) 2-structures in Fig. \ref{Hardy}.

The weak values of the localized rank-2 projector corresponding to individual particle states are,
\begin{equation}
|L^+|_w = |L^+R^-|_w + |L^+L^-|_w = 0,
\end{equation}
\begin{equation}
|R^+|_w = |R^+R^-|_w + |R^+L^-|_w = 1,
\end{equation}
\begin{equation}
|L^-|_w = |L^+L^-|_w + |R^+L^-|_w = 1,
\end{equation}
and
\begin{equation}
|R^-|_w = |L^+R^-|_w + |R^+R^-|_w = 0.
\end{equation}
These correspond to the sum of the counterparticles at each location.

In the ABL interpretation, these weak values force us to conclude that the electron and positron both took the inner arms of their respective interferometers, and yet they both reached the detectors, and so must have passed without annihilating, which is the original Hardy paradox.

The $N$-structure picture for this scenario is shown in Fig. \ref{Hardy}.  Weak measurements of the rank-2 projectors corresponding to the individual arms probe the corresponding counterparticles, while weak measurements of the rank-1 projectors corresponding to arm-products between different systems probe the corresponding 2-structures, with the pointer receiving a quasi-classical impulse in each case.
 
In this description, we still have a positron and electron on the inner arms where they would annihilate, but because they are not truly isolated particles, but rather the component ends of different 2-structures, the annihilation is prevented.  We are essentially positing a physical rule that both ends of these 2-structures would need to meet simultaneously to produce an annihilation, and this rule provides us with an elegant resolution of the paradox.

\section{Discussion}

We hope that we have clearly conveyed the top-down weak reality as a retrocausal model with quasi-classical particles that move on well-defined trajectories during the time between pre- and post-selection, and that we have convinced the reader that this model provides an intuitively useful particle-based picture of the underlying physics of unmeasured systems.

The examples in the SI further emphasize the generality and versatility of the weak reality model, which may, we believe, form the foundation for a new retrocausal formulation of quantum physics based on particles rather than waves.  We plan to develop the model further to see where this insight leads.

\textbf{Acknowledgments}:--- This research was supported (in part) by the Fetzer-Franklin Fund of the John E. Fetzer Memorial Trust.  This research was supported by grant number (FQXi-RFP-CPW-2006) from the Foundational Questions Institute and Fetzer Franklin Fund, a donor advised fund of Silicon Valley Community Foundation. E.C. was supported by the Israeli Innovation authority (grants 70002 and 73795), by the Pazy foundation and by the Quantum Science and Technology Program of the Israeli Council of Higher Education. J.T. and Y.A. thank the Federico and Elvia Faggin Foundation for support.
\pagebreak
\bibliographystyle{pnas-new}
\bibliography{Counterparticles_Biblio}

\end{document}


\title{Supporting Information: Quantum reality with negative-mass particles}

\author{Mordecai Waegell$^{{a,b}}$, Eliahu Cohen$^{c,d}$, Avshalom Elitzur$^{a,d}$, Jeff Tollaksen$^{a,b}$, Yakir Aharonov$^{a,b,d,e}$}

\affiliation{\footnotesize{$^a$Institute for Quantum Studies, Chapman University, 1 University Dr., Orange, CA 92866, USA\\$^b$Schmid College of Science and Technology, Chapman University, 450 N Center St., Orange, CA 92866, USA\\$^c$Faculty of Engineering and the Institute of Nanotechnology and Advanced Materials, Bar Ilan University, Ramat Gan 5290002, Israel\\$^d$Iyar, The Israeli Institute for Advanced Research, POB 651 Zichron Ya'akov 3095303, Israel\\$^e$School of Physics and Astronomy, Tel Aviv University, Tel Aviv, Israel}}

\begin{abstract}\footnotesize{Quantum Physics, Weak Values, Quantum Paradoxes, Time-Symmetry, Quantum Measurement}
\end{abstract}

\maketitle
\tableofcontents

\section{Strong and Weak Measurement}  \label{Measure}

We begin with a review of the quantum measurement theory which gives rise to both the ABL rule in the limit of strong projective measurement, and the weak value in the limit of weak measurements.  As we will see, the weak value becomes encoded into the pointer wavefunction of a measurement device when the translation induced by the coupling Hamiltonian is so small relative to the width of the pointer wavefunction that the different terms interfere --- a \emph{weak measurement} --- and the ensemble is both pre- and post-selected.

We can model a general measurement by considering the position wavefunction of the pointer system as it is moved along some ruler by the measurement interaction.  Let us consider the case that the pointer wavefunction is a Gaussian of width $\epsilon$, and the ruler tick marks that indicate orthogonal states of the measured system are a distance $d$ apart.  The usual case of a strong projective measurement is the case that $\epsilon \ll d$, which means that the pointer has a very narrow peak, always centered on a tick mark of the ruler.  The other extreme case where $\epsilon \gg d$ is the regime of weak measurements, in which case each Gaussian is broadly spread around its ruler mark, and may overlap and interfere with Gaussian terms centered on other marks.  If we consider a pre- and post-selected ensemble of weak measurements, this interference results in a Gaussian that is centered at the weak value (to first order in $d/\epsilon$).

As an example, let us consider a measurement of a single spin-1/2 particle using a continuous Gaussian pointer.  The impulsive coupling Hamiltonian between the spin and the pointer is given by,
\begin{equation}
\hat{H} = g(t)\sigma_z \hat{P}_p,
\end{equation}
where $\sigma_z = |0\rangle\langle 0 | - |1\rangle\langle 1 |$ is the Pauli operator of the qubit, $\hat{P}_p$ is the momentum operator of the pointer system, and $g(t)$ is the coupling strength.  We will assume that the interaction lasts for only a very brief period $\tau$, and that during this time, we can neglect the systems' other unitary evolution.  This $g_o = \int_0^\tau g(t)dt$ is the relevant coupling parameter.  The general initial state of the qubit is,
\begin{equation}
|\phi_0\rangle = a|0\rangle + b|1\rangle,
\end{equation}
which is expressed in the $\sigma_z$ basis, with $|a|^2 + |b|^2 = 1$, and the initial wavefunction of the pointer is
\begin{equation}
\psi(x) = (\epsilon^2 \pi)^{-1/4}e^{-x^2/2\epsilon^2}
\end{equation}
The system and pointer begin in the product state,
\begin{equation}
|\Psi_0\rangle = |\phi_0\rangle \int_{-\infty}^\infty \psi(x) |x\rangle dx.
\end{equation}
After the interaction, the two systems are in the entangled state,
\begin{equation}
|\Psi\rangle = e^{-i g_0\sigma_z P_p /\hbar}|\Psi_0\rangle
\end{equation}
\begin{equation}
    =\left(a|0\rangle  e^{-i g_0 P_p / 2} + b|1\rangle e^{i g_0 P_p / 2}\right)\int_{-\infty}^\infty \psi(x) |x\rangle dx \nonumber
\end{equation}
\begin{equation}
=\int_{-\infty}^\infty dx|x\rangle [ a|0\rangle\psi(x-d)+ b|1\rangle\psi(x+d) ], \nonumber
\end{equation}
with $d=g_0$, which sets the scale of the ruler.

Now, suppose we project the qubit onto a general normalized state $|\phi_f\rangle = \alpha|0\rangle + \beta|1\rangle$ (the post-selection).

First we consider the strong measurement limit $\epsilon \rightarrow 0$, where $\psi(x) \rightarrow \sqrt{\delta(x)}$, and the renormalized wavefunction is,
\begin{equation}
\psi_f^s(x) = \big(\alpha^* a \sqrt{\delta(x-d)} + \beta^* b \sqrt{\delta(x+d)}\big)/\sqrt{|\alpha a|^2 + |\beta b|^2}.
\end{equation}
Taking the probability to find the pointer at positions $x = \pm d$, we obtain the ABL probability rule for finding eigenstate $|j\rangle$ ($j\in\{0,1\}$) when an intermediate strong projective measurement of $\sigma_z$ is made,
\begin{equation}
P_{\textrm{ABL}}(|j\rangle\langle j| = 1 \space \; \vert \; \phi_f, \phi_0, \sigma_z) = \frac{|\langle \phi_f | j \rangle \langle j | \phi_0\rangle|^2}{ \sum_{k\in\{0,1\}} |\langle \phi_f | k \rangle \langle k | \phi_0\rangle|^2}.
\end{equation}
For a large PPS ensemble, the \emph{conditional expectation value} of the intermediate measurement of $\sigma_z$,
\begin{equation}
\langle \hat{\sigma_z} \rangle_{ABL}  = \sum_{j} \lambda_j P_{\textrm{ABL}}(|j\rangle\langle j| = 1 \space \; \vert \; \phi_f, \phi_0, \sigma_z)\label{Cond}
\end{equation}
\begin{equation}
= P_{\textrm{ABL}}(|0\rangle\langle 0| = 1 \space \; \vert \; \phi_f, \phi_0, \sigma_z) - P_{\textrm{ABL}}(|1\rangle\langle 1| = 1 \space \; \vert \; \phi_f, \phi_0, \sigma_z)  \nonumber
\end{equation}
\begin{equation}
   =\frac{|\alpha a|^2 - |\beta b|^2}{|\alpha a|^2 + |\beta b|^2}, \nonumber
\end{equation}
where $\lambda_j$ is the eigenvalue corresponding to eigenstate $|j\rangle$.

Next we consider the weak measurement regime where $\epsilon \gg d$, the two $\psi$ terms explicitly interfere, and the renormalized pointer wavefunction is,
\begin{equation}
\psi_f^w(x) = \big(\alpha^* a \psi(x-d) + \beta^* b \psi(x+d)\big)/\langle \phi_f | \phi_0 \rangle   \label{psi}
\end{equation}
\begin{equation}
    = (\epsilon^2 \pi)^{-1/4}(\alpha^* ae^{-(x-d)^2/2\epsilon^2}  + \beta^* b e^{-(x+d)^2/2\epsilon^2})/\langle \phi_f | \phi_0 \rangle \nonumber
\end{equation}
\begin{equation}
 \approx  (\epsilon^2 \pi)^{-1/4}e^{-(x^2+d^2)/2\epsilon^2} \bigg(\frac{ \alpha^* a  (1+xd/\epsilon^2)  + \beta^* b (1-xd/\epsilon^2)}{\alpha^* a + \beta^* b}\bigg)\nonumber
\end{equation}
\begin{equation}
  =(\epsilon^2 \pi)^{-1/4}e^{-(x^2+d^2)/2\epsilon^2}\left(1  + \frac{\alpha^* a - \beta^* b}{\alpha^* a + \beta^* b}(xd/\epsilon^2) \right)   \nonumber
\end{equation}
\begin{equation}
\approx  (\epsilon^2 \pi)^{-1/4}\exp{\left(-\frac{1}{2\epsilon^2}\left(x - d\left[\frac{\alpha^* a - \beta^* b}{\alpha^* a + \beta^* b}\right]\right)^2\right)}  \nonumber
\end{equation}
\begin{equation}
  = \psi(x- d(\sigma_z)_w), \nonumber
\end{equation}

where,
\begin{equation}
(\sigma_z)_w \equiv \frac{\langle \phi_f |\sigma_z | \phi_0\rangle}{\langle \phi_f | \phi_0\rangle} = \frac{\alpha^* a - \beta^* b}{\alpha^* a + \beta^* b},
\end{equation}
is the weak value of $\sigma_z$ given the pre-selection $|\phi_0\rangle$ and the post-selection $\langle \phi_f |$.  Note that the weak value has emerged from the interference of two Gaussian terms with complex coefficients coming from both the pre- and post-selection.

Let us contrast the two limits:  In the strong case, the final pointer function is a superposition of two delta functions at $x = \pm d$, meaning one obtains a definite eigenvalue shot-by-shot, and the mean value $d\langle \hat{\sigma_z} \rangle_{ABL}$ emerges from a PPS ensemble.  In the weak case, the final pointer function is a broad Gaussian, meaning one effectively obtains only noise shot-by-shot, and the mean value $d(\sigma_z)_w$ emerges from a PPS ensemble.

As we have seen, the weak value emerges due to interference of the pointer, mediated (or steered) by entanglement with the measured system.  Now, the initial pointer state had $\langle x \rangle_0=0$ and $\langle p \rangle_0=0$, and it is straightforward to check that the final pointer state has $\langle x \rangle_f=d\textrm{Re}[(\sigma_z)_w]$ and $\langle p \rangle_f \approx d\hbar\textrm{Im}[(\sigma_z)_w)]/\epsilon^2$, and thus we see that the real part of the weak value is proportional to the shift in the pointer position, while the imaginary part is proportional to the shift in the pointer momentum (to first order).  This gives us a physical interpretation of the complex weak value.

The derivation above can be naturally generalized to measure any observable $\hat{A}$ on any physical system using the same pointer, and the Hamiltonian, $\hat{H} = g(t)\hat{A}\hat{P}_p$ to measure the weak value, $A_w$.


In the limit that $\epsilon \gg d$, the weak value is the dominant effect on the pointer wavefunction, and as one takes the limit that $d/\epsilon \rightarrow 0$, the weak value is always encoded in the pointer wavefunction, right down to the limit that there is no interaction at all.  The premise of the weak value interpretation is that all of these weak values are physically existent properties of the system, whether or not we choose to weakly measure them.  This is reminiscent of how the electric field is defined at each location by considering the limit that the magnitude of a test charge placed at that location goes to zero, whether or not we actually measure the force the field exerts on a charge at that location.  In this picture, it is the weak values which describe how nature behaves when we are not looking, and the usual eigenvalues which describe how it behaves when we are.

\section{Resolving Paradoxes}

Here we explore a number of PPS-scenarios and PPS-paradoxes and construct the relevant weak values.  We then work out the top-down counterparticle description in the weak reality, and discuss the resolution of the corresponding paradoxes (if any) in the ABL interpretation.

\subsection{The Disappearing and Reappearing Particle Paradox}\label{Dis}
The 3-box paradox can be generalized to a time-dependent case in an interesting way \cite{aharonov2017case}.  Suppose that a particle is pre-selected in the state $|\psi\rangle = (|1\rangle + \sqrt{2}|2\rangle)/\sqrt{3}$ at time $t=0$, and post-selected in the state $|\phi\rangle=(|1\rangle - i\sqrt{2}|3\rangle)/\sqrt{3}$ at $t=2$.  Boxes 2 and 3 share a wall that allows tunneling, while box 1 is isolated from both of them.  Due to the tunneling, the amplitude of the wavefunction will oscillate back and forth between between the two boxes according to the unitary evolution matrix,
\begin{equation}
U(\Delta t) = \left(
  \begin{array}{ccc}
    1 & 0 & 0 \\
    0 & \cos(\pi \Delta t/4) & i\sin(\pi \Delta t/4) \\
    0 & i\sin(\pi \Delta t/4) & \cos(\pi \Delta t/4) \\
  \end{array}
\right).
\end{equation}
Using this expression we can propagate $|\psi\rangle$ forward and $|\phi\rangle$ backward to an intermediate time $t$.  This gives $|\psi_t\rangle = (|1\rangle + \sqrt{2}\cos(\pi t / 4)|2\rangle + i\sqrt{2}\sin(\pi t / 4)|3\rangle)/\sqrt{3} $ and $|\phi_t\rangle = (|1\rangle + \sqrt{2}\sin(\pi t / 4)|2\rangle - i\sqrt{2}\cos(\pi t / 4)|3\rangle)/\sqrt{3} $, where we call $|\phi_t\rangle$ the \emph{destiny} vector.  With these we can compute the weak value of each projector at time $t$ as, $|1_t|_w = 1$, $|2_t|_w = \sin(\pi t/2)$, and $|3_t|_w = -\sin(\pi t/2)$.

Consider the weak value $|\Pi^{23}_t|_w = |2_t|_w + |3_t|_w = 0$.  According to the  ABL interpretation, this should mean that there is never a particle in either of boxes 2 or 3, and thus the particle must be in box 1 at all times.  However, at $t=1$, we have $|\Pi^{13}_1|_w = |1_1|_w + |3_1|_w = 0$, meaning the particle must be in box 2, at $t=3$ we have $|\Pi^{12}_3|_w = |1_3|_w + |2_3|_w = 0$, meaning the particle must be in box 3.  At these two times, we have recovered exactly the original 3-box paradox.  The added subtlety is that the particle seems to definitely be in box 1 at all times, but then at $t=1$ it seems paradoxically to also be in box 2, even though there is no tunneling between boxes 1 and 2, and likewise it seems to also be in box 3 at $t=3$.

The counterparticle ontology has a single positive particle in box 1 with probability 1, however the situation in boxes 2 and 3 is more complicated. At $t=0$ there probability 1/2 to find a positive particle in box 2 and a negative one in box 2, and also probability 1/2 to find them reversed - leading to weak value of 0.  As time evolves, each of these starting configurations of counterparticles has some probability to flip due to tunneling, such that the total probability of finding the first configuration is $P_{+-} = [1-\sin(\pi t/2)]/2$ and the total probability to find the second is $P_{-+} = [1+\sin(\pi t/2)]/2$.

As this happens, the negative real particles can mask the positive real particles so that sometimes it looks as though a given box is empty, and in time it looks as though the positive-real particle in boxes 2 or 3 gradually disappears and then gradually reappears as a negative-real particle, only to reverse course, vanishing and reappearing as a positive particle before the entire process repeats.  

Note that the total probability to find a negative particle in boxes 2 or 3 is always 1, as is the probability find a positive particle in boxes 2 or 3.  Note also, that provided the pre-selected state at $t=0$ is $|\psi\rangle$, we can choose the post-selection $|\phi_t\rangle$ at any time $t$ and the ontology is the same.

\subsection{The Case of the Hollow Atoms}

In the following thought experiment we will analyze another case of the 3-box paradox using tripartite system composed of two protons (hereby denoted by $p_1$,$p_2$) and one electron ($e$) superposed over 3 boxes. The electron is assumed to bind to either proton if they are left in the same box, thus forming a hydrogen atom. As we shall see, upon a particular choice of pre- and post-selected states, the weak reality will tell us a peculiar story.

The system is prepared in the state
\begin{equation}
|\psi\rangle=|1^{p_1}\rangle|1^e\rangle|3^{p_2}\rangle + |2^{p_1}\rangle|2^e\rangle|3^{p_2}\rangle +|2^{p_1}\rangle|3^e\rangle|3^{p_2}\rangle,
\end{equation}

and post-selected in the state:
\begin{equation}
|\phi\rangle=|1^{p_1}\rangle|1^e\rangle|3^{p_2}\rangle + |2^{p_1}\rangle|2^e\rangle|3^{p_2}\rangle -|2^{p_1}\rangle|3^e\rangle|3^{p_2}\rangle,
\end{equation}
Both the pre- and post-selected states therefore represent the case where there is a hydrogen atom superposed over the three boxes and there is always a ``spectator'' proton, which is sometimes the separable proton in the third box and sometimes is the entangled proton in the second box.

The nonzero weak values of the rank-1 projectors are,

$|1^{p_1}1^{e}|_w = 1$, $|2^{p_1}2^{e}|_w = 1$, $|2^{p_1}3^{e}|_w = -1$ $|3^{p_2}|_w = 1$, and the rank-2 projector weak values for $p_1$ are,
$|1^{p_1}|_w = |1^{p_1}2^{e}|_w = 1$,  $|2^{p_1}|_w = |2^{p_1}2^{e}|_w + |2^{p_1}3^{e}|_w = 0$, 
and for $e$ they are
$|1^{e}|_w = |1^{p_1}1^{e}|_w = 1$, 
$|2^{e}|_w = |2^{p_1}2^{e}|_w = 1$, 
and $|3^{e}|_w = |2^{p_1}3^{e}|_w = -1$.

In terms of single-particle weak values, the second term in the pre- and post-selected states indicates the effective presence of a positive electron and a positive proton within the second box. However, the third term implies the effective presence of a proton counter-particle (i.e. the weak value of the corresponding projector is equal to $-1$). Therefore, in the weak reality, the proton particle and counter-particle effectively cancel and we seem to have in total just one electron within the second box. However, the two-particle weak value of the projector onto an electron-proton {\it pair} (henceforth an ``atom'') within the second box is $|2^{p_1}2^{e}|_w = 1$.  We can thus interpret the ABL paradox here as implying that the electron in box 2 is bound to an empty nucleus, thus forming a `Hollow Atom.'

The top-down 2-structures and counterparticles of the weak value ontology of this case has a positive 2-structure with a $p_1$ and an $e$ in box 1, a positive 2-structure with a $p_1$ and an $e$ both in box 2, a negative 2-structure with a $p_1$ in box 2 and an $e$ in box 3, and single positive $p_2$ in box 3.  This explains the weak values, and there is no paradox.

Remarkably, the ABL paradox here may suggest novel atomic structures in the weak value ontology composed of counterparticles and $N$-structures.


\subsection{The 4-Box Paradox}

The 4-box paradox is much less discussed, so we formally introduce it here before moving on to several better-known examples.  The projector weak values in the 4-box paradox are $(1,1,1,-1)/2$, or $(\Pi_1)_w = (\Pi_2)_w = (\Pi_3)_w = 1/2$, and $(\Pi_4)_w = -1/2$.  To see the paradox, we must consider the three coarse-grained dichotomic bases $\mathcal{B}_1 = \big(\Pi_1 + \Pi_2, \Pi_3 + \Pi_4\big)$, $\mathcal{B}_2 = \big(\Pi_1 + \Pi_3, \Pi_2 + \Pi_4\big)$, and $\mathcal{B}_3 = \big(\Pi_2 + \Pi_3, \Pi_1 + \Pi_4\big)$, which all have weak values 0 and 1.  These three bases tell us that in the ABL interpretation, the particle must be in boxes (1 or 2), and also (1 or 3), and also (2 or 3), giving us a logical contradiction.

\subsection{The (Original) Quantum Cheshire Cat Paradox}
The quantum Cheshire Cat paradox \cite{aharonov2013quantum, denkmayr2014observation} is given by a composite 4-level system composed of the spin and path degrees of freedom of a neutron in a Mach-Zehnder interferometer.  The pre-selected entangled state inside the interferometer is,
\begin{equation}
|\psi\rangle = (|\uparrow\rangle|L\rangle + |\uparrow\rangle|R\rangle + |\downarrow\rangle|L\rangle - |\downarrow\rangle|R\rangle)/2,
\end{equation}
and the post-selected product state is,
\begin{equation}
|\phi\rangle  = (|\uparrow\rangle|L\rangle + |\uparrow\rangle|R\rangle + |\downarrow\rangle|L\rangle + |\downarrow\rangle|R\rangle)/2.
\end{equation}
 Using compact notation, the weak values of the projectors onto these four basis states are,\\
 $|\uparrow L|_w = |\uparrow R|_w = |\downarrow L|_w = 1/2$, and $|\downarrow R|_w=-1/2$, from which we see that this a 4-box paradox.  We also find the weak values of the six rank-2 projectors projectors,
\begin{equation}
|\uparrow|_w = |\uparrow L|_w + |\uparrow R|_w = 1,
\end{equation}
\begin{equation}
|\downarrow|_w = |\downarrow L|_w + |\downarrow R|_w = 0,
\end{equation}
\begin{equation}
|L|_w = |\uparrow L|_w + |\downarrow L|_w = 1,
\end{equation}
\begin{equation}
|R|_w = |\uparrow R|_w + |\downarrow R|_w = 0,
\end{equation}
\begin{equation}
|\circlearrowleft|_w =  |\downarrow L|_w + |\uparrow R|_w= 1,
\end{equation}
\begin{equation}
|\circlearrowright|_w =|\uparrow L|_w + |\downarrow R|_w = 0,
\end{equation}
which pair up into complete dichotomic measurement bases.

Following the ABL-weak-value correspondence rule, the paradox here is that $|\uparrow|_w = 1$ implies that the particle has spin up, and $|L|_w = 1$ implies that it is in the left arm of the interferometer, and $|\circlearrowleft|_w = 1$ implies that a spin up particle must take the right arm, while a spin down particle must take the left arm, and these three statements are mutually contradictory.

A fantastical interpretation of this contradiction is that the neutron's spin becomes disembodied from its mass, allowing the up spin to travel the right arm, while the spinless mass travels the left \cite{aharonov2015current}.  Indeed, experiments seem to show that in the left arm there will be evidence of massive particles where no spin is detected, and in the right arm there is no evidence of massive particles where a spin is detected.  This disembodiment effect is called the Quantum Cheshire Cat in reference to a disembodied grin without a cat from \emph{Alice in Wonderland}.

The simplest resolution of this paradox in the weak value ontology has a probability 1/2 for a single spin-up particle on the left path, and a probability 1/2 to have a spin-up on the right path along with a negative spin-down, and positive spin-down on the right.  Thus a mass detector always finds something on the left, and finds an average of zero on the left since the particle is negative half the time.  And a spin detector finds an average of zero on the left, since the spin is up and down with equal probability, while on the right a spin up is always detected, since a negative spin-down couples in the same way as a positive spin-up due to its opposite charge.

This ontology provides a clear explanation for the experimental observations related to the Quantum Cheshire Cat, without the paradoxical spatial separation of the neutron's spin and mass.

As an aside, we can think of these rank-1 projectors as 2-structures, but since the spin is an internal property of the particle, it is a single localized object.

\subsection{The Quantum Pigeonhole Paradox}

The quantum pigeonhole paradox \cite{aharonov2016quantum, waegell2018contextuality, waegell2017confined} uses three 2-level systems (pigeons in of two boxes) all pre-selected in the state $|\psi\rangle = (|L\rangle + |R\rangle)/\sqrt{2}$ and post-selected in the state $|\phi\rangle = (|L\rangle + i|R\rangle)/\sqrt{2}$, where $|L\rangle$ and $|R\rangle$ are two boxes (pigeonholes).  For each pigeon, the weak value of the projector into the left box is $|L|_w = (1+i)/2$, and for the right box it is $|R|_w = (1-i)/2$.  Because these are independent systems we know that the weak value of the tensor product is also the product of the individual weak values.  This allows us to deduce that for any two of the three pigeons, $|LL|_w = i/2$, $|LR|_w = 1/2$, $|RL|_w = 1/2$, $|RR|_w = -i/2$, and from these we can construct the weak values $|S|_w = |LL|_w+|RR|_w = 0$ for the projector onto both pigeons being in the same box, and $|O|_w = |LR|_w + |RL|_w = 1$ for the two pigeons being in opposite boxes.

Following the ABL-weak-value-correspondence rule, the paradox is that $|O|_w^{12} = 1$ implies that pigeons 1 and 2 are in opposite boxes, $|O|_w^{13} = 1$ implies that pigeons 1 and 3 are in opposite boxes, and $|O|_w^{23} = 1$ implies that pigeons 2 and 3 are in opposite boxes, and these three statements are mutually contradictory.  In fact, they violate the \emph{pigeonhole principle} which states that if three (classical) pigeons are placed into two boxes, then one of the boxes must have two or more pigeons in it.

The 4-box paradox is also the key to the Quantum Pigeonhole Effect.  To see this we must consider the projectors onto states of all three particles, $|LLL|_w = (-1+i)/4$, $|LLR|_w = (1+i)/4$, $|LRL|_w = (1+i)/4$, $|LRR|_w = (1-i)/4$, $|RLL|_w = (1+i)/4$, $|RLR|_w = (1-i)/4$, $|RRL|_w = (1-i)/4$, $|RRR|_w = (-1-i)/4$.  Then we construct the coarse-grained projector weak values, $|LLL|_w + |RRR|_w = -1/2$ and $|LLR|_w + |RRL|_w = |LRL|_w + |RLR|_w = |RLL|_w + |LRR|_w = 1/2$, which gives us the 4-box paradox.

This paradox can also be seen from the conditional correlations that appear between each pair of pigeons, even though they never interact in this PPS.  The conditional correlation is defined using the conditional expectation value formula of Eq. \ref{Cond}, $\langle \hat{\sigma}_z \rangle_{ABL} = 0$ for each pigeon and $\langle \hat{\sigma}^1_z \hat{\sigma}^2_z \rangle_{ABL} = -1$ for each pair of pigeons, with $\hat{\sigma}_z \equiv |L\rangle\langle L| - |R\rangle\langle R|$.  As a result, the conditional covariance $\textrm{cov}_{ABL}(\hat{\sigma}^1_z,\hat{\sigma}^2_z) = \langle \hat{\sigma}^1_z \hat{\sigma}^2_z \rangle_{ABL} - \langle \hat{\sigma}^1_z \rangle_{ABL}\langle \hat{\sigma}^2_z \rangle_{ABL} = -1$, which suggests that two causally disconnected systems are strongly reproducibly correlated, which is a logical contradiction.

In the weak value interpretation, each pigeon appears in the left box or right box with probability 1/2, and independently there is an imaginary positive-negative pair that appears with probability 1/2, with the positive imaginary particle on the left, and the negative on on the right.  The products then arise as the products of three such independent distributions, and explain all of the weak values - without any paradoxical correlations between the three independent systems.

\subsection{The All-or-Nothing Paradox}

Next we explore the All-or-Nothing paradox on $N$ 3-level systems (see also \cite{aharonov2018completely}).  Ignoring normalization, the pre-selection is $|\psi\rangle = \bigotimes^N_{i=1}(|1\rangle - |2\rangle) + \bigotimes^N_{i=1}|3\rangle$, and the post-selection is $|\phi\rangle = \bigotimes^N_{i=1}(|1\rangle + |2\rangle) + \bigotimes^N_{i=1}|3\rangle$.

For example, let $N=2$.  There are nine projectors in the joint product basis of the two systems, with weak values, $|11|_w = |22|_w = |33|_w = 1$, $|12|_w = |21|_w = -1$, and $|13|_w = |23|_w  = |31|_w = |32|_w  = 0$.

This can be seen as an extended 3-box paradox, where a +1, a -1, and four 0 weak values have been added to the set, and the paradox obtains in more combinations.  The All-or-Nothing paradox is based on another observation about this set, which starts with constructing the coarse-grained projector weak values of the individual systems,
 \begin{equation}
 |1^1|_w = |11|_w + |12|_w + |13|_w = 0, 
\end{equation}
 \begin{equation}
|2^1|_w = |21|_w + |22|_w + |23|_w= 0,
\end{equation}
 \begin{equation}
|3^1|_w = |31|_w + |32|_w + |33|_w= 1,
\end{equation}
 \begin{equation}
 |1^2|_w = |11|_w + |21|_w + |31|_w= 0,
\end{equation}
 \begin{equation}
|2^2|_w = |12|_w + |22|_w + |32|_w = 0,
\end{equation}
and
 \begin{equation}
|3^2|_w = |13|_w + |23|_w + |33|_w = 1.
\end{equation}
In the ABL interpretation, these weak values show that neither system can be in states $\Pi_1$ or $\Pi_2$, and thus they must always be in $\Pi_3$.  But this contradicts the nonzero weak values $|11|_w = |22|_w = 1$ which both appear in dichotomic coarse-grained bases, and indicate that the system must also always be in orthogonal states $\Pi_1$ and $\Pi_2$.  This is the All-or-Nothing paradox for boxes 1 and 2 -- only joint weak values of $\Pi_1$ or $\Pi_2$ for all $N$ systems can have nonzero weak values, whereas any joint weak value of $\Pi_1$ or $\Pi_2$ for $N-1$ or fewer systems has zero weak value.  Specifically, if a strong projective measurement is made in the product basis $\{|i,j\rangle\}$ during the interval between the pre- and post-selection, then the ABL formula gives a probability of 1/5 to find the system in any of the states, |11|,|12|, |21|, |22|, or |33|, whereas if a projective measurement is made on only one of the two systems, then the ABL probability is 1 to find the system in |3|.  Thus, if one hides the third box, it seems very literally that boxes 1 and 2 are either both empty, or contain all $N$ systems.

In the weak value ontology, there are three positive 2-structures with both particles in the same box, for boxes 1, 2 and 3, and for there are 2-negative 2-structures, each with one particle in box 1 and the other in box 2.  This means that in box 1 there are a total of two positive particles and 2 negative particles, which hide each other, and likewise in box 2. The four 2-structures for boxes 1 and 2 produces nonzero values for joint measurements.  This explain all of the weak values for this scenario, without any paradoxes.

It is easy to check that for larger values of $N$ the results are similar, with zero weak value for any rank-1 projector that includes box 3, and so the All-or-Nothing property is general for all $N$.

\subsection{The Hermit Particle}

The Hermit Particle is not actually a logical PPS paradox.  Instead, it is a case where the logic of the ABL interpretation is plausible, but results in a very counterintuitive situation.  Consider the set of projector weak values $(\Pi_1)_w = \delta$,  $(\Pi_2)_w = -1 + \delta$, $(\Pi_3)_w = 1 - \delta$, $(\Pi_4)_w = 1 - \delta$, with real positive $\delta \ll 1$.  In the ABL interpretation the particle is always found in the first state, no matter how small $\delta$ is made.  To see this, consider the two coarse-grained dichotomic bases with weak values 0 and 1, $\mathcal{B}_1 = (\Pi_1 + \Pi_3, \Pi_2 + \Pi_4)$ and $\mathcal{B}_2 = (\Pi_1 + \Pi_4, \Pi_2 + \Pi_3)$.  This shows that the system must be in states ($\Pi_1$ or $\Pi_3$) and ($\Pi_1$ or $\Pi_4$), from which we logically conclude it must be in state $\Pi_1$.  However, if we perform a projective measurement of the dichotomic basis $\mathcal{B}_3 = (\Pi_1,  \Pi_2 +\Pi_3 + \Pi_4)$ during the interval between the pre- and post-selection, then the ABL formula give a probability on the order of $\delta^2 \ll 1$.  Thus according to the ABL interpretation, the particle is always located where it is least likely to be found -- and thus a hermit.

The Hermit Particle can be generalized to a $d$-level system ($d>4$) by adding additional projector weak values $(\Pi_i)_w = 1 - \delta$, and resetting $(\Pi_2)_w = (-1 + \delta)(d-3)$.  In the limit $\delta \rightarrow 0$, the case of the Hermit Particle reduces to the extended 3-box paradox.

In the weak value interpretation, the there is a particle in box 1 with probability $\delta$, and with probability $1-\delta$ there is a negative particle in box 2, and a positive particle in box 3 along with another in box 4.

\subsection{The Energy Teleportation Paradox}

The Energy Teleportation Paradox is not a logical PPS paradox.  Instead the paradox is that energy seems to be transferred from one system to another without ever passing through the space between them -- thus teleported \cite{elouard2019spooky, waegell2020energy}.  This case is closely related to Hardy's paradox and interaction free measurement \cite{elitzur1993quantum}.  

A particle with average energy $E_p$ is sent through a Mach-Zehnder interferometer (MZ), where it may strike a quantum object in one of the arms.  If the particle strikes the object, it is always absorbed or deflected, and does not reach the second beam splitter of the MZ.  The object starts in a low energy eigenstate, $|0^o\rangle = (|\textrm{in}\rangle + |\textrm{out}\rangle )/\sqrt{2}$ of its confining potential with energy eigenvalue $E_0$, which is a superposition of being inside and outside arm I of the MZ.  The energy eigenstate $|1^o\rangle = (|\textrm{in}\rangle - |\textrm{out}\rangle )/\sqrt{2}$ has energy eigenvalue $E_1 > E_0$.

Once the particle inside the MZ has reached the object, the two enter an entangled state because of the nonzero probability for the particle to scatter off of the object.  We will make this state our (unnormalized) pre-selection, $|\psi\rangle = |\textrm{I}^p\rangle |\textrm{out}^o\rangle + |\textrm{II}^p\rangle |\textrm{in}^o\rangle + |\textrm{II}^p\rangle |\textrm{out}^o\rangle$.  

Evolving this through the second beam splitter, we obtain $|\psi'\rangle = 2|\textrm{br}^p\rangle |\textrm{out}^o\rangle + |\textrm{br}^p\rangle |\textrm{in}^o\rangle - |\textrm{dk}^p\rangle |\textrm{in}^o\rangle$, where `br' is the bright port, and `dk' is the dark port of the MZ.  From $|\psi'\rangle$ we can see that projecting the particle onto the dark port also projects the object inside the MZ.  Thus we have detected that the object is inside arm I of the MZ, but the particle must have taken taken arm II, since the object in arm I would have scattered it out of the MZ and it would never have reached the dark port.  Because the particle detects the object without ever going near it, we have an interaction-free measurement.  Furthermore, the object has been left in the state $|\textrm{in}^o\rangle$, which is superposition of energy eigenstates with average energy $(E_0 + E_1)/2$.  This means that on average, the particle has delivered $\Delta E = (E_1 - E_0)/2$ to the object, without ever going near it, and thus the energy appears to have been teleported.

We now take the post-selection $|\phi'\rangle = |\textrm{dk}^p\rangle |\textrm{in}^o\rangle$, which we can retropropagate back through the second beam splitter to obtain $|\phi\rangle = \big( |\textrm{I}^p - |\textrm{II}^p\big)\rangle |\textrm{in}^o\rangle/\sqrt{2}$.  Now, with the unprimed PPS, we consider the weak values, $|\textrm{I}^p 0^o|_w = -1/2$, $|\textrm{I}^p 1^o|_w = 1/2$, $|\textrm{II}^p 0^o|_w = 1$, and $|\textrm{II}^p 1^o|_w = 0$.  Considering the coarse-grained dichotomic basis, $\mathcal{B} = (|\textrm{I}^p 0^o| + |\textrm{I}^p 1^o|, |\textrm{II}^p 0^o| + |\textrm{II}^p 1^o|)$, we see that in the ABL interpretation, the particle definitely takes path II of the MZ.  

The projector weak values of the energy eigenstates of the object are $|0^o|_w = |\textrm{I}^p 0^o|_w + |\textrm{II}^p 0^o|_w = 1/2$ and $|1^o|_w = |\textrm{I}^p 1^o|_w + |\textrm{II}^p 1^o|_w = 1/2$, and thus the weak value of the object's energy is $(E_0 + E_1)/2$.  The extra energy $\Delta E$ was delivered by the particle due to an energy-conserving local interaction Hamiltonian in arm I.

In the weak value interpretation, there is a positive 2-structure with a particle on arm II with average energy $E_p$ and an object in the ground state of energy $E_0$ with probability 1.  This 2-structure corresponds to zero energy transfer between the particle and object.  Furthermore, with probability 1/2 there is a positive-negative pair of 2-structures; a positive 2-structure with a particle on arm I with energy $E_p - (E_1 - E_0)$ and an object in the excited state of energy $E_1$, and a negative 2-structure with a particle on arm I with energy $-E_p$ and the object in the ground state of energy $-E_0$.  Then the average energy of the particle is $\bar{E}_p = E_p + [E_p - E_1 + E_0 - E_p]/2 = E_p - (E_1 -E_0)/2$, and the average energy of the object is $\bar{E}_o = E_0 + [E_1 - E_0]/2 = (E_0 + E_1)/2$.  And of particular interest, on arm II there is no exchange between the particle and object, while on arm I the average energy of the particle and object are $\bar{E}_{p,I} =[E_p - E_1 + E_0 - E_p]/2 = - \Delta E$, and $\bar{E}_o = \Delta E$, respectively, and so the particle effectively gains this energy while emitting a packet of negative energy which is absorbed into the particle during post-selection.

If we insist that all counterparticles must exist during the entire interval between the pre-selection before the MZ to the post-selection at the dark port, then we see that the positive and negative particle in arm I were already present before the interaction with the object, and this is where their total energies became different from zero.

Thus, the counterparticle model gives us a satisfying resolution to the Energy Teleportation Paradox, where instead of a nonlocal transfer, the object gets the energy from a local interaction with a particle in arm I.  In order for this effect to obtain, the incident particle must have an energy uncertainty $\sigma \gg \Delta E$, so that the particle can deliver the energy without significantly reducing the visibility of interference at the second beam splitter.

\subsection{Three entangled 2-position systems}

\begin{figure}[h]
    \centering
    \caption{The subpairs and corresponding 2-structures for of the 3-party system.  All of the edges of the corresponding 3-structure can be read off of these three diagrams.}\label{2webs}
    \begin{tabular}{ccccc}
    \includegraphics[width=1.7in]{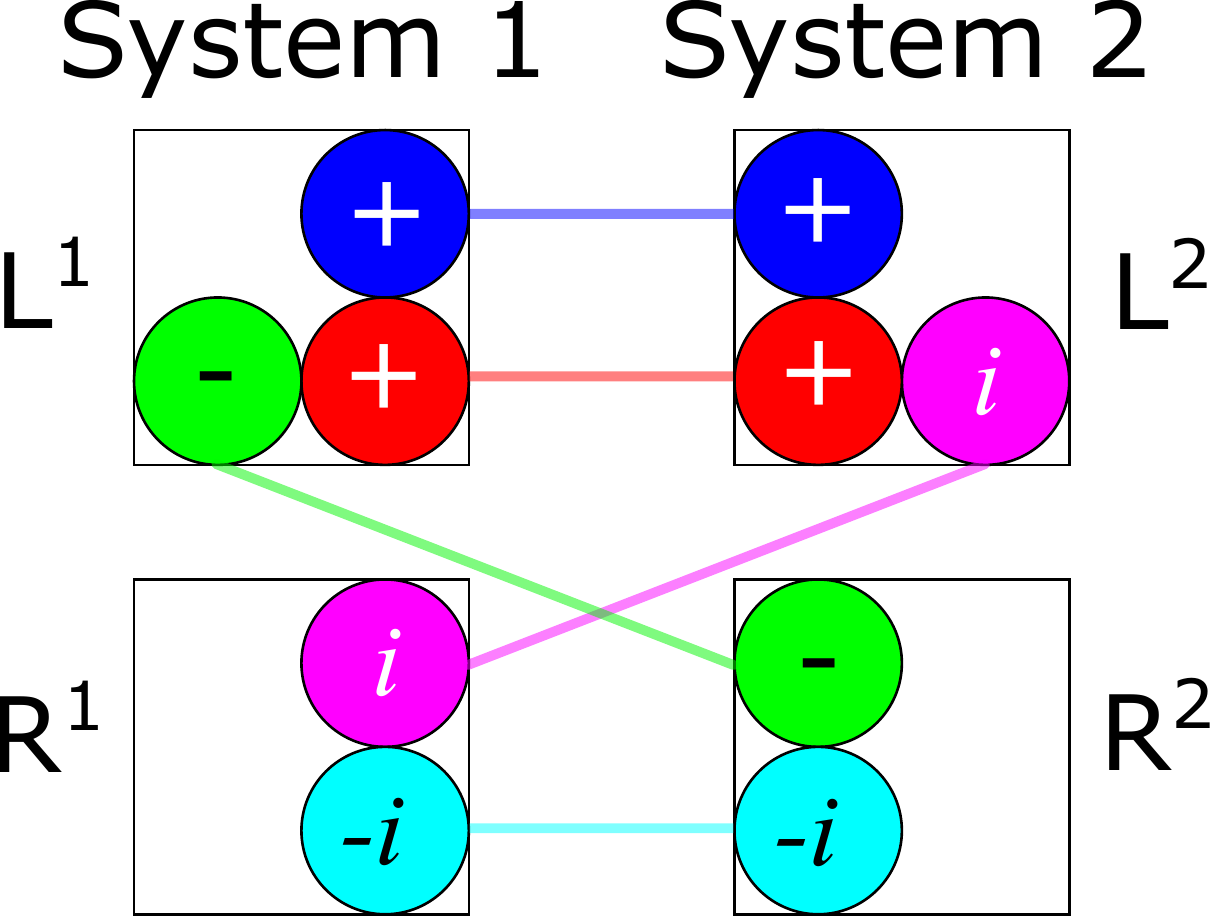}     & &  \includegraphics[width=1.7in]{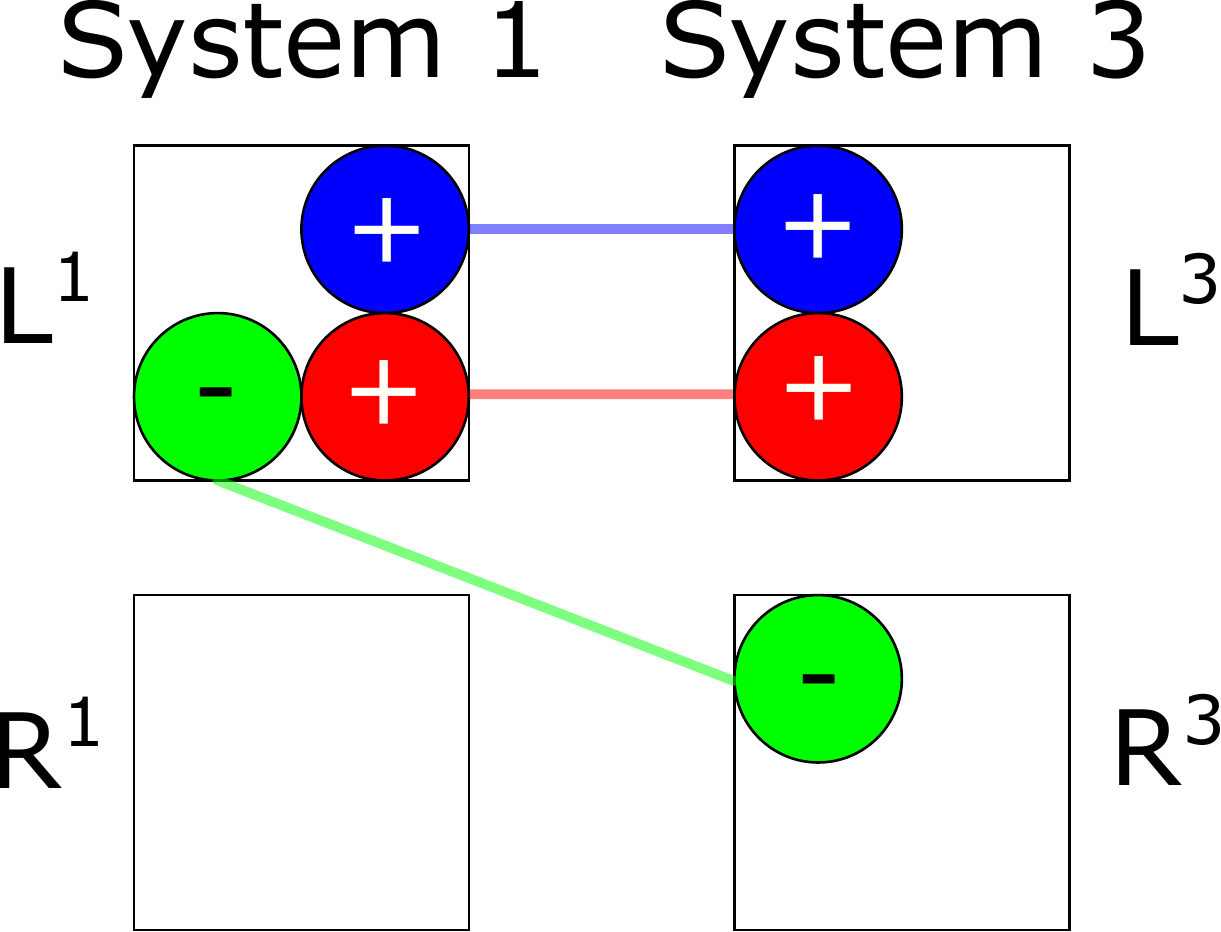} & &  \includegraphics[width=1.7in]{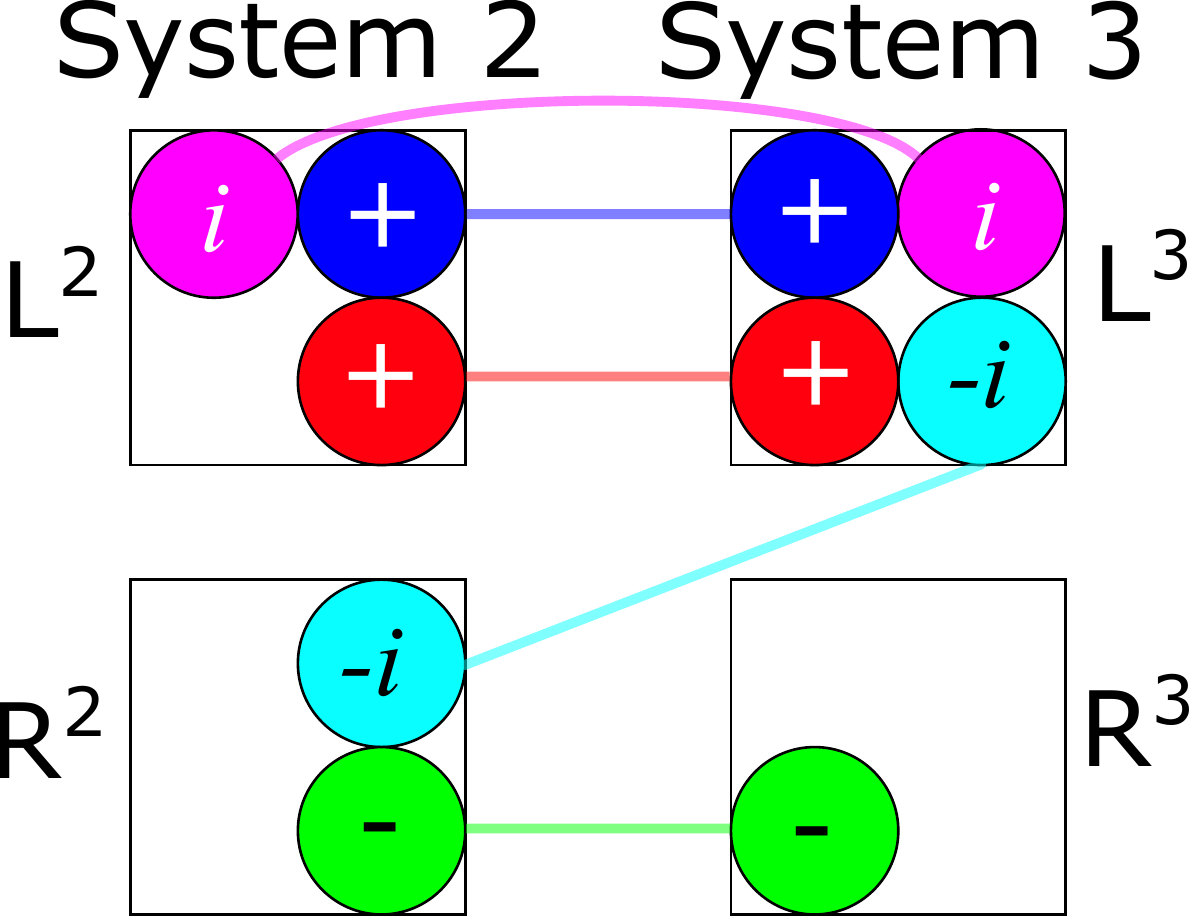} \\
    \end{tabular}
\end{figure}

Consider three 2-position quantum systems, each with an eigenbasis $(|L\rangle, |R\rangle)$ which are prepared (pre-selected) in the entangled state,
\begin{equation}
   |\psi\rangle =  \big(2|LLL\rangle - |LRR\rangle +i|RLL\rangle -i|RRL\rangle\big)/\sqrt{7},
\end{equation}
and post-selected in the product state 
\begin{equation}
     |\phi\rangle =  \big(|L\rangle +|R\rangle\big)\big(|L\rangle +|R\rangle\big)\big(|L\rangle +|R\rangle\big)/\sqrt{8}.
\end{equation}
The weak values of the eight rank-1 projector are $|LLL|_w = 2$, $|LRR|_w = -1$, $|RLL|_w = i$, $|RRL|_w = -i$, and $|LLR|_w = |LRL|_w = |RLR|_w = |RRR|_w = 0$, which correspond to five (nonzero) 3-structures.

\begin{figure}[b]
    \centering
    \includegraphics[width=4in]{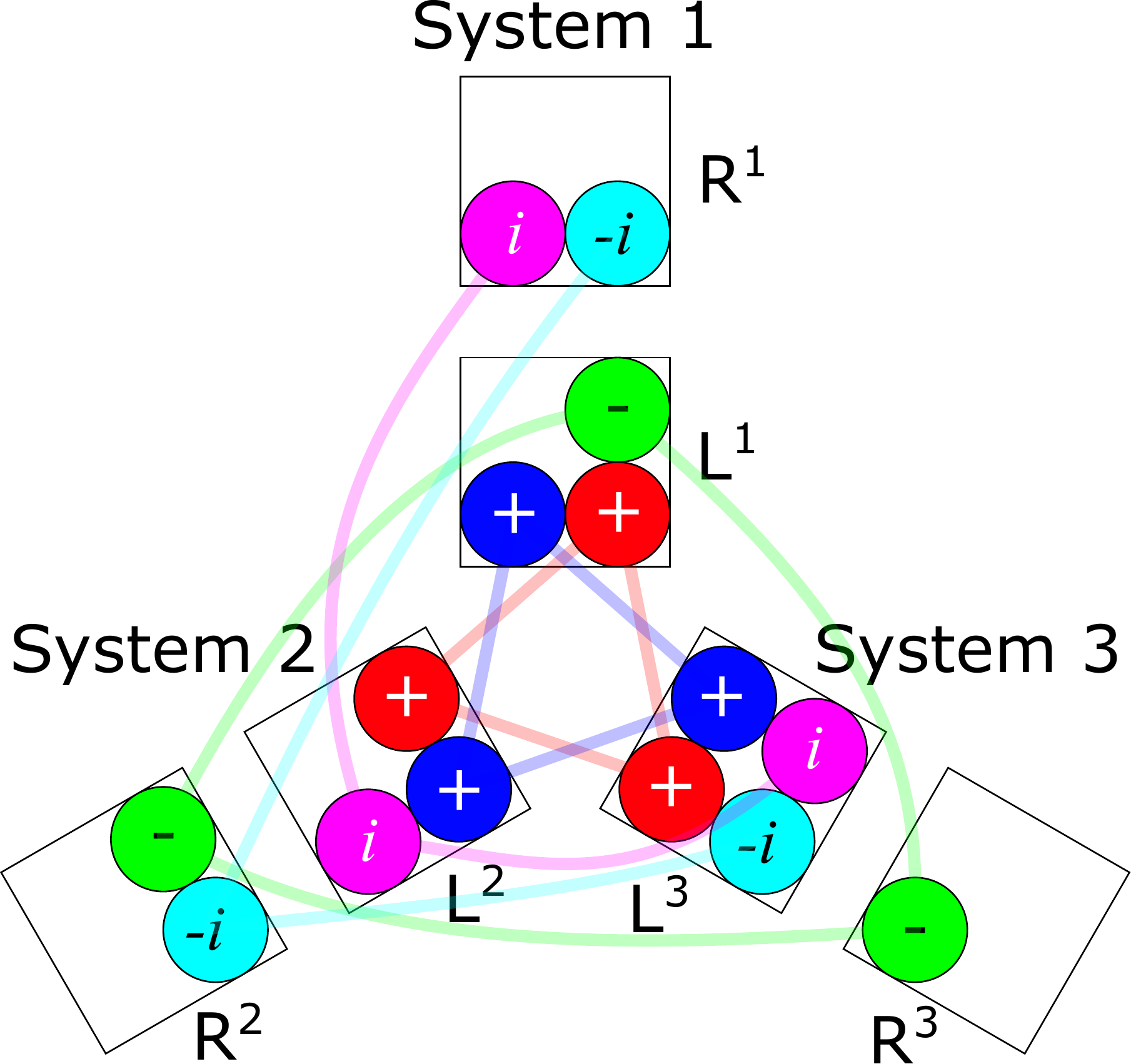}
    \caption{The five 3-structures, each a different color, for the entangled pre-selection $|\psi\rangle =  \big(2|LLL\rangle - |LRR\rangle +i|RLL\rangle -i|RRL\rangle\big)/\sqrt{7}$ and post-selection $|\phi\rangle =  \big(|L\rangle +|R\rangle\big)\big(|L\rangle +|R\rangle\big)\big(|L\rangle +|R\rangle\big)/\sqrt{8}$, and their corresponding counterparticles, with each web assigned a different color.  The 2-structures for any 2 subsystems are just the subgraphs of the graphs shown here, which can be verified by considering Fig. \protect\ref{2webs}
    - and the 1-structures are just the counterparticles themselves.}
    \label{3web}
\end{figure}

To find the edges of these 3-structures we will need to consider the 2-structures corresponding to each pair of fundamental subsystems.  Summing over system 3 we find the weak values for the product projectors onto systems 1 and 2,
\begin{equation}
  |L^1L^2|_w = |LLL|_w + |LLR|_w = 2,
\end{equation}
\begin{equation}
  |L^1R^2|_w = |LRL|_w + |LRR|_w = -1,
\end{equation}
\begin{equation}
  |R^1L^2|_w = |RLL|_w + |RLR|_w = i,
\end{equation}
\begin{equation}
  |R^1R^2|_w = |RRL|_w + |RRR|_w = -i,
\end{equation}
summing over system 2 we find the weak values for the product projectors onto systems 1 and 3,
\begin{equation}
  |L^1L^3|_w = |LLL|_w + |LRL|_w =2,
\end{equation}
\begin{equation}
  |L^1R^3|_w = |LLR|_w + |LRR|_w = -1,
\end{equation}
\begin{equation}
  |R^1L^3|_w = |RLL|_w + |RRL|_w = 0,
\end{equation}
\begin{equation}
  |R^1R^3|_w = |RLR|_w + |RRR|_w = 0,
\end{equation}
and summing over system 1 we find the weak values for the product projectors onto systems 2 and 3,
\begin{equation}
  |L^2L^3|_w = |LLL|_w + |RLL|_w =2+i,
\end{equation}
\begin{equation}
  |L^2R^3|_w = |LLR|_w + |RLR|_w = 0,
\end{equation}
\begin{equation}
  |R^2L^3|_w = |LRL|_w + |RRL|_w = -i,
\end{equation}
\begin{equation}
  |R^2R^3|_w = |LRR|_w + |RRR|_w = -1.
\end{equation}
The configurations of 2-structures for these subpairs are shown in Fig \ref{2webs}.

The five 3-structures for this PPS are shown in Fig. \ref{3web}.  Note that for this example the three real 3-structures are fully connected, while the two imaginary 3-structures are not.

This example also allows us to consider measurements of lower-rank projectors of an entangled system.  To measure an $N$-structure, the pointer must be somehow coupled to the rank-1 projector, which is a product of the values of all $N$ systems at $N$ particular locations, and thus the pointer must somehow interact at all $N$ locations during the PPS interval.  In all cases, the end result is that the pointer has received a quasi-classical impulsive shift corresponding to the weak value.  For projectors corresponding to $n<N$ systems and locations, the pointer only needs to couple to the $n$ sites, and the shift is determined using the $n$-structure subgraph of the $N$-structure corresponding to those systems.  Note that a physical weak measurement inherently disturbs both systems, and so there is always some mixture of the purely impulsive shift due to the counterparticles and higher-order entanglement effects, which are negligible in the weak regime.

\section{Discussion}

\subsection{The Counterparticle Representation of All Observables}

All of the cases we have examine so far considered a single fine-grained basis for a given PPS, along with various coarse-grainings of that same basis, since this is where the PPS paradoxes originate, but the weak values are defined for every observable of the system, and the counterparticle model should provide a single consistent description of all of these weak values.  We show that this can be done by choosing a cardinal set of observables, and rather than a counterparticle being in just one state of the system, it is simultaneously in one eigenstate of each cardinal observable.

For a single cardinal direction $n$, we have a set of distributions of counterparticle configurations $\{C_{ni}\}$, each of which occurs with probability $P_{ni}$.  Each $C_{ni}$ is a vector with the same dimension as the Hilbert space of the system, and each component is a complex integer corresponding to a definite set of real and/or imaginary counterparticles.  The general distribution is obtained by taking the Cartesian product of all such cardinal sets, multiplying the corresponding probabilities, and the corresponding counterparticle sets.  Note that each quasi-classical 2-level particle is in a definite state of all three observables, even though none of them commute.  The weak value of just one of the cardinal projectors is obtained by summing over all values of the projectors belonging to other cardinal observables.
 
 It will be useful to introduce a pseudo-density matrix for the PPS which we call the \textit{upsidedown state}, $\breve{\rho} \equiv |\psi\rangle\langle  \phi | / \langle \phi |\psi\rangle$, which satisfies $\breve{\rho}^2 = \breve{\rho}$ and Tr$\breve{\rho} = 1$, but is clearly non-Hermitian.  The utility of the upsidedown state is that we now have $A_w =$Tr$\breve{\rho}A$, in analogy to the usual expectation value $\langle A \rangle =$Tr$\rho A$.

Now, in a 2-level system, the cardinal set is $(\sigma_x, \sigma_y, \sigma_z)$, which we can see by decomposing the upside-down state for a general PPS as, 
\begin{equation}
\breve{\rho} \equiv \frac{|\psi\rangle\langle\phi|}{\langle\phi|\psi\rangle} =  \big(I + (\sigma_x)_w \sigma_x +  (\sigma_y)_w\sigma_y + (\sigma_z)_w\sigma_z \big)/2 = (I+\vec{w}\cdot\vec{\sigma})/2,
\end{equation}
where $\vec{w}$ is the vector of cardinal weak values.  The weak value of an observable $A$ is then,
\begin{equation}
A_w = \textrm{Tr}(\breve{\rho}A) = \big[\textrm{Tr}(A) + (\sigma_x)_w \textrm{Tr}(\sigma_xA) +  (\sigma_y)_w \textrm{Tr}(\sigma_yA) + (\sigma_z)_w\textrm{Tr}(\sigma_zA) \big]/2 = A_0 +\vec{w}\cdot\vec{A},
\end{equation}
 where $\vec{A}$ is the vector of normalized expectation values $A_x = \textrm{Tr}(\sigma_xA)/2$,  $A_y = \textrm{Tr}(\sigma_yA)/2$, $A_z = \textrm{Tr}(\sigma_zA)/2$, and $A_0 = \textrm{Tr}(A)/2$.  For the case of a the Pauli observable in the direction of unit vector $\hat{n}$, this reduces to the simple form,
\begin{equation}
( \sigma_{\hat{n}} )_w  = \vec{w}\cdot\hat{n}.
\end{equation}
Thus we can obtain the weak values along any direction $\hat{n}$ simply by knowing the cardinal weak values, and we can still use the same set of particles and counterparticles in their combined eigenstates to obtain the weak value in any direction.

Of course, we could rotate the coordinate system and find a new representation in terms of counterparticles for the new cardinal observables.  This is yet another symmetry of the counterparticle model, since the descriptions of the same physical system may call for a completely different counterparticle representation in the different coordinate system.

As a simple example, consider a single pigeon from the quantum pigeonhole paradox, with pre-selection $|\psi\rangle  = |L\rangle + |R\rangle$ and post-selection $|\phi\rangle  = |L\rangle + i|R\rangle$ for each 2-level system.  Recalling that $\sigma_z = |L\rangle\langle L| - |R\rangle\langle R|$, we also have $\sigma_x|\psi\rangle = |\psi\rangle$, and $\sigma_y|\phi\rangle = |\phi\rangle$, we can see that the weak vector is $\vec{w} = (1, 1, i)$.  We thus have upside-down state $\breve{\rho} = (I + \sigma_x +  \sigma_y + i\sigma_z )/2$.  For $x$ and $y$ the simplest configurations both have a single positive particle in $\Pi_{+x}$ and $\Pi_{+y}$ ($C_{x1} = [1, 0]$, and $C_{y1} = [1, 0]$), respectively, with probability 1.  For $z$, the simplest case we can use has, with probability $P_{z1} = 1/2$, a single positive particle in $\Pi_{+z}$ ($C_{z1} = [1, 0])$, and with probability $P_{z2} = 1/2$ there is a positive imaginary particle in $\Pi_{+z}$ and positive real particle plus a negative imaginary particle in $\Pi_{-z}$ ($C_{z2} = [1+i, i]$).  The Cartesian product of these three distributions has, with probability $P_1 = 1/2$, a positive counterparticle in joint state $\Pi_{+x, +y, +z}$ ($C_1 = [1, 0, 0, 0, 0, 0, 0, 0, 0]$), and with probability $P_2 = 1/2$ a positive imaginary particle in $\Pi_{+x, +y, +z}$ and positive real particle plus a negative imaginary particle in $\Pi_{+x, +y, -z}$ ($C_2 = [i, 1-i, 0, 0, 0, 0, 0, 0, 0]$), where the configurations $C_j$ now include counterparticles of all eight joint types in canonical order.

To make the situation slightly more interesting, we rotate the coordinate system so that the direction of the post-selection changes from $\hat{y}$ to $(\hat{y} + \hat{z})/\sqrt{2}$ on the Bloch sphere, which is still in the plane perpendicular to the direction of the pre-selection $\hat{x}$.  Formally, we have $|\phi'\rangle = \cos(\pi/8)|L\rangle + i\sin(\pi/8)|R\rangle$ and $|\psi'\rangle = |\psi\rangle$, which results in the cardinal weak vector $\vec{w} = \big(1, (1-i)/\sqrt{2}, (1+i)/\sqrt{2}\big)$.  The same upside down state is now represented as $\breve{\rho} = \Big(I + \sigma_x + \frac{1-i}{\sqrt{2}}\sigma_y + \frac{1+i}{\sqrt{2}}\sigma_z\Big)/2$ in the new coordinates, from which we obtain $(\Pi^\pm_y)_w = (1 \pm (1-i)/\sqrt{2})$ and $(\Pi^\pm_z)_w = (1 \pm (1+i)/\sqrt{2})$. 

As before we have $\Pi_{+x}$ with probability 1, but for the $y$ direction we now have $P_{y1} = \frac{1}{2\sqrt{2}}$, $P_{y2} = \frac{1}{2}$, and $P_{y3} = \frac{1}{2} - \frac{1}{2\sqrt{2}}$, with corresponding configurations $C_{y1} = [1-i, i]$, $C_{y2} = [1, 0]$, and $C_{y3} = [0, 1]$, and for the $z$ direction we have $P_{z1} = \frac{1}{2\sqrt{2}}$, $P_{z2} = \frac{1}{2}$, and $P_{z3} = \frac{1}{2} - \frac{1}{2\sqrt{2}}$, with corresponding configurations $C_{z1} = [1+i, -i]$, $C_{z2} = [1, 0]$, and $C_{z3} = [0, 1]$.

The joint distribution then has nine configurations, each with a probability and a set of counterparticles obtained by taking the products of the three individual sets.  For example, the probability to obtain the first configuration is $P_1 = (1)(\frac{1}{2\sqrt{2}})(\frac{1}{2\sqrt{2}}) = \frac{1}{8}$, and the set of counterparticles in the joint state $\Pi_{+x, +y, +z}$ in that configuration is $(1)(1-i)(1+i) = 2$.  The full set of probabilities and configurations is given in Fig. \ref{Complete9B}

 \begin{figure}[h]
\centering
\caption{The probabilities and simplified counterparticle representation of the PPS with upside-down state $\breve{\rho} = \Big(I + \sigma_x + \frac{1-i}{\sqrt{2}}\sigma_y + \frac{1+i}{\sqrt{2}}\sigma_z\Big)/2$.  The marginal weak value of $\Pi^\pm_i$ is given by summing over the corresponding columns.} \label{Complete9B}
\begin{tabular}{c|c|c|c|c|c|c|c|c}
$P$ & $(\Pi_x^+,\Pi_y^+,\Pi_z^+)$ & $(\Pi_x^+,\Pi_y^+,\Pi_z^-)$ & $(\Pi_x^+,\Pi_y^-,\Pi_z^+)$ & $(\Pi_x^+,\Pi_y^-,\Pi_z^-)$
& $(\Pi_x^-,\Pi_y^+,\Pi_z^+)$ & $(\Pi_x^-,\Pi_y^+,\Pi_z^-)$ & $(\Pi_x^-,\Pi_y^-,\Pi_z^+)$ & $(\Pi_x^-,\Pi_y^-,\Pi_z^-)$ \\
\hline
$\frac{1}{8}$ & $2$ & $-1 - i$  & $-1+i$ & $1$ & $0$ & $0$ & $0$ & $0$\\
$\frac{1}{4\sqrt{2}}$ & $1-i$ & $0$  & $i$ & $0$ & $0$ & $0$ & $0$ & $0$\\
$\frac{1}{4\sqrt{2}} - \frac{1}{8}$ & $0$ & $1-i$  & $0$ & $i$ & $0$ & $0$ & $0$ & $0$\\
$\frac{1}{4\sqrt{2}}$ & $1+i$ & $-i$  & $0$ & $0$ & $0$ & $0$ & $0$ & $0$\\
$\frac{1}{4}$ & $1$ & $0$  & $0$ & $0$ & $0$ & $0$ & $0$ & $0$\\
$\frac{1}{4} - \frac{1}{4\sqrt{2}}$ & $0$ & $1$  & $0$ & $0$ & $0$ & $0$ & $0$ & $0$\\
$\frac{1}{4\sqrt{2}} - \frac{1}{8}$ & $0$ & $0$  & $1+i$ & $-i$ & $0$ & $0$ & $0$ & $0$\\
$\frac{1}{4} - \frac{1}{4\sqrt{2}}$ & $0$ & $0$  & $1$ & $0$ & $0$ & $0$ & $0$ & $0$\\
$\frac{3}{8} - \frac{1}{2\sqrt{2}}$ & $0$ & $0$  & $0$ & $1$ & $0$ & $0$ & $0$ & $0$\\
\end{tabular}
\end{figure}

Now, when the 2-level system is the spin of a particle, it is straighforward to imagine a single particle which somehow simultaneously has the three properties $\sigma_x = \sigma_y = \sigma_z = +1$, but if the system is a spatial superposition, then the interpretation is more subtle.  If the $\Pi^{\pm}_z$ are orthogonal spatial states, like the arms of an MZ, then $\Pi^{\pm}_x$ and $\Pi^{\pm}_y$ are spatial superpositions, so how can a particle with a single trajectory be in a simultaneous state of all three?  The simplest answer seems to be that the spatial basis is the one that tells us where the particle is actually located, since each projectors can be measured at only one location.  The other states are internal properties of the particle, which are only revealed by a measurement which couples to the system at both locations (measurements of say $\sigma_x$ or $\sigma_y$), similar to measuring $N$-structures.

Finally, this type of cardinal representation can be found in all dimensions by expanding the upside-down state into the set of $d$-dimensional generalized Gell-Mann matrices.  This works just as in the $d=2$ case above, because the Gell-Mann matrices are all traceless, as is the product of any two of them, and the trace of their squares are always 2.  There are $d^2-1$ different Hermitian Gell-Mann matrices, which combined with the identity span the space of all observables of the system (with real coefficients, and the space of all operators with complex coefficients).  This means the general expanded form of the upside-down state is,
\begin{equation}
\breve{\rho} = \frac{1}{d}I + \frac{1}{2}\sum_{i=1}^{d^2-1} (g_i)_w g_i = \frac{1}{d}I + \frac{1}{2} \vec{w}_d \cdot \vec{g},
\end{equation}
and thus a complete counterparticle representation can always be constructed in this way, with $\vec{g}$ a vector of the $d^2-1$ generalized Gell-Mann matrices, and  $\vec{w}_d$ the vector of their weak values.  Likewise a general observable can be said to point in a particular direction $\hat{n}_d$ in the $(d^2-1)$-dimensional real space of cardinal matrices, and we can expand a general Gell-Mann observable as $g_{\hat{n}_d} = \hat{n}_d \cdot \vec{g}$, and we again obtain,
\begin{equation}
( g_{\hat{n}_d} )_w  = \vec{w}_d\cdot\hat{n}_d,
\end{equation}
which shows that this cardinal counterparticle representation defines the weak values of all observables of the system.

Finally for dimensions $d = 2^N$, a similar expansion can be constructed using the observables of the $N$-qubit Pauli group, which are all tensor products, and so this representation may be more convenient for some applications.


\subsection{Intermediate Interaction Strength}

The counterparticle model is a good physical approximation only in the limit of weak measurements.  The weak limit is the first order approximation for small $d/\epsilon$.  We expect that a more elaborate model can be devised in which higher order terms also have a physical interpretation, and become increasingly relevant as the interaction strength increases.  In the projective measurement limit, this would require interpreting an infinite number of terms, and more importantly, this is the limit where the interaction induces a post-selection, and so as the interaction strength is slowly increased from zero, at first there may be higher order objects in the counterparticle model (we have not yet explored this), but then this entire picture starts to give way, to be replaced by a new collapse event.  This raises a more fundamental questions about the range of physical situations this model can describe.  

In particular, consider a system that undergoes frequent periodic measurement interactions, each of small, but not insignificant strength.  This system is unlikely to collapse due to any one measurement, and instead its continuous weak measurement readout will wander back and forth, only eventually collapsing to an eigenvalue after many measurements \cite{garcia2017past}.  Then it will tend to remain there for a little while due to an effect called \emph{Zeno pinning}, but eventually its Hamiltonian dynamics will set it back to wandering, and then to another random collapse.

Now, consider a long period of wandering during which the system never actually collapses to either eigenvalue, which can be more readily accomplished by alternating kicks in complementary measurement bases \cite{garcia2016probing}.  It has been shown in numerical simulations that due to the random kicks, the state of a system rapidly becomes independent of its previous states, even if it never fully collapses.  The simulations also show that the continuous weak measurement readout likewise becomes rapidly independent of the future states of the system.  This means that the intermediate (nearly weak) values that are being continuously read out simply do not have a projective pre-selection or a projective post-selection.  Instead, the pre- and post-selection occurred gradually, over the course of enough random kicks to screen a small time interval off from both its past and future.  To give a rough idea of how this could work, suppose that the measurement events are indexed by $i$, and the state is screen off after $n$ such events.  Then the physical readout at event $i$ will be centered on roughly the weak value, using the some function of the physical states at events $i-1$, $i-2$,..., $i-n$, as the pre-selection, and some function of $i+1$,  $i+2$,..., $i+n$ as the post-selection.  The trouble with this is that we cannot actually know the physical state at each event, as we do in the case of projective measurements, and this seems to be the price we pay for the counterparticle model to apply to physical situations like this.  Nevertheless, nature knows the PPS, even when it cannot be experimentally observed.

\bibliographystyle{pnas-new}
\bibliography{Counterparticles_Biblio}